\documentclass[a4paper,11pt]{article}
\pdfoutput=1 

\usepackage{jheppub} 

\usepackage[T1]{fontenc} 
\usepackage[all]{xy}
\usepackage{epsfig,psfrag}
\usepackage{graphicx}
\usepackage{rotating}
\usepackage{ragged2e}
\usepackage{float}
\usepackage{placeins}
\usepackage{appendix}

\makeatletter
\AtBeginDocument{%
  \expandafter\renewcommand\expandafter\subsection\expandafter
    {\expandafter\@fb@secFB\subsection}%
  \newcommand\@fb@secFB{\FloatBarrier
    \gdef\@fb@afterHHook{\@fb@topbarrier \gdef\@fb@afterHHook{}}}%
  \g@addto@macro\@afterheading{\@fb@afterHHook}%
  \gdef\@fb@afterHHook{}%
}
\makeatother

\def\be{\begin{equation}}
\def\ee{\end{equation}}
\def\zz{z,\hat{z}}

\title{\boldmath D-brane superpotentials, $SU(2)$ Ooguri-Vafa invariants and TypeII/F -theory duality}

\author{Xiao-Tian Jiang,}
\author[1]{Fu-Zhong Yang\note{Corresponding author.}}

\affiliation{University of Chinese Academy of Sciences, \\No.19(A) Yuquan Road, Shijingshan District, Beijing, P.R.China}

\emailAdd{fzyang@ucas.ac.cn}

\abstract{The phase transitions are studied for the D-brane systems with multiple open-string moduli in terms of toric geometry: between the parallel D-brane phase corresponding to the Coulomb branch and the coincident phase corresponding to the Higgs branch. The two separated D-branes on compact Calabi-Yau 3-fold coincide developing the geometric singularity in the corresponding F-theory Calabi-Yau 4-fold in terms of TypeII/F theory duality, and the enhancement of gauge group $U(1)\times U(1)\rightarrow SU(2)$ in terms of gauge theory.
For several D-brane system with various closed-string moduli, using the mirror symmetry and the typeII/F theory duality the A-model superpotentials are obtained from the B-model side for the two phases, and the Ooguri-Vafa invariants are extracted from the A-model superpotential.
We find the discrete $Z_2$ symmetry of superpotentials in the two parallel D-branes phase of all the models which is a signal of decoupling of the parallel topological D-branes.
Furthermore, the \emph{$U(1)$ Ooguri-Vafa invariants} for one of the two parallel D-branes are the same as the invariants for the D-brane system with only one D-brane.
However they are different from the \emph{$SU(2)$ Ooguri-Vafa invariants} corresponding to the two D-branes coincident phase.
We present these Ooguri-Vafa invariants for two phases with figures to observe the difference, and find that in two phases the points for invariants form a wave-packet respectively. The wave-packets for coincident phase are higher and wider than ones for parallel phase which means more complicate spectrum structure.
This is an evidence of the phase transition between the Coulomb branch and the Higgs branch.}

\begin{document}
\maketitle
\flushbottom

\section{Introduction}
\label{sec:intro}

In topological string theory, $N=2$ mirror symmetry gives equivalence between two geometrically distinct Calabi-Yau compactifications of string theory, namely A-model which is featured by Kahler moduli and B-model which is characterized by complex structure moduli.
When D-branes are included, it gives rise to the $N=1$ special geometry breaking the supersymmetry to $N=1$ and leading to the open-closed mirror symmetry.
Similar to the prepotential in the $N=2$ supersymmetric situation, the exact non-perturbative holomorphic superpotential that determines the F-term of low-energy effective theory and the string vacuum structure, is available from a B-model computation.
The techniques of localization \cite{AV.00,AKV.02,A.05}, topological vertex and direct integration relate to $N=1$ special geometry \cite{LMW.02.1,LMW.02.2} are developed for dealing with the computation on non-compact Calabi-Yau manifolds. For more complicate compact Calabi-Yau manifolds, other methods can be found in prior works, such as mixed Hodge structure variation, Gauss-Manin connection \cite{JS.08,JS.09}, the blow-up method \cite{GHKK.09} and the GKZ-generalized hypergeometric system for open-closed sectors \cite{AHJMMS.09,AHJMMS.10,XY.14.1,XY.14.2,CXY.14,ZY.15,ZY.16}.
In addition the expression of instanton expansion of superpotential on the A-model side encodes the number of BPS states and mathematically it corresponds to the Ooguri-Vafa invariants, which are related to the open Gromov-Witten invariants and can be interpreted as counting holomorphic disks \cite{AV.00}.
The algebraic geometric treatments of the open Gromov-Witten theory of a smooth projective variety X depend on the moduli stack $\overline{\cal{M}}_{g,h,n} (X, C,\beta)$ which parameterizes the stable maps from genus $g$ curves with $h$ boundaries and $n$ marked points to $(X, C)$ with image class of the curve $\beta \in H_{2}(X, C, Z)$ and with image class of boundaries of the curve being in $ H_{1}(C, Z)$, where the submanifold $C$ of $X$ is wrapped by the D-branes from the physical viewpoint \cite{FLT.12}. When $ h =0$, the moduli stack $\overline{\cal{M}}_{g,h,n} (X, C, \beta) $ just is the Kontsevich's moduli stack $\overline{\cal{M}}_{g,n} (X, \beta)$ which is a proper Deligne-Mumford stack with projective coarse moduli space \cite{BM.96,KM.94}. In particular, if $X$ is a point, then $\overline{\cal{M}}_{g,n} (*)$ just is the Deligne-Mumford moduli stack $\overline{\cal{M}}_{g,n}$ of stable curves \cite{DM.69,V.89}.

So far, most of computations of superpotentials for the D-brane system with the compact target space are contributed by branes which are described by only one open-string deformation.
However, the superpotentials in Type-II string theory have a dual description in the F-theory as the flux superpotentials equalizing the open and closed parameters as the complex structure moduli of the 4-fold for F-theory compactification \cite{M.01}.
This suggests a way to obtain the superpotentials of Type-II string theory compactified on Calabi-Yau 3-fold from the computation on a Calabi-Yau 4-fold compactification of F-theory \cite{AHJMMS.09,JMW.09}.
We apply it in the two D-branes system taking the advantage that the information of D-branes are totally encoded into the combinatoric data of the 4-fold for F-theory. Remarkably, when D-branes approach to each other, the enhancement of the gauge symmetry on the D-brane worldvolume is equivalent to the appearance of the singularity on the dual 4-fold of the F-theory. A connection between the singularities and non-perturbative enhanced gauge symmetries has been established in \cite{BV.95}. And a similar effect in a exceptional $N=2$ compactification with a $N=4$ characteristic spectrum has been investigated in \cite{A.96}. For studying the coincident phase, we construct the corresponding singular 4-folds in terms of toric variety and the superpotentials of coincident D-brane are obtained using the type II/F-theory duality.

The focus of this paper push the study of D-brane system to the more complicate cases which is the D-branes system involving multiple open-string moduli and the parallel/coincident D-branes phase of the system corresponding to the Coulomb branch and Higgs branch of the non-Abelian gauge theory on the D-branes' worldvolume respectively.
In three exact model, namely the mirror quintic and the hypersurface in the weighted projective space $P(1,1,2,2,2)$ and $P(1,1,1,3,3)$, we study the two D-branes system and calculate the superpotentials near the large radius limit point for the first time.
The discrete $Z_2$ symmetry group of the two parallel D-branes is interpreted as the Weyl symmetry of the non-perturbative $SU(2)$ gauge theory, which acts on the positions of the individual D-branes in the Coulomb branch of the $SU(2)$ gauge theory. We found the Weyl symmetry in the calculations of parallel cases meeting the expectation in terms of the instanton expansion of the superpotential.
Furthermore, it's remarkable that the $SU(2)$ gauge symmetry is the enhancement of the $U(1)\times U(1)$ gauge theory on the worldvolume of the separate two D-branes due to the twins approach to each other.
In the coincident D-branes phase of the open-closed string moduli space, the coincidence of two D-branes restricts the degree of freedom of the open-closed string deformations. In our model the independent couple of open-string moduli are reduced to one and it seems can be identified with the single D-brane system geometrically. However, the superpotentials of the coincident D-branes and the superpotential for single D-brane system are different in our calculations. The $U(1)$ and $SU(2)$ Ooguri-Vafa invariants are extracted from the superpotentials. And the invariants for two phases are presented on one figure for comparing which displays absolutely different BPS state spectrum. It can be understand as a signal of phase transition.

The organization of this paper is as follows.
In section 2, we introduce the background and formalism in three parts. In the first part, we recall the D-brane superpotential in the Type II string theory, the Gukov-Vafa-Witten (GVW) superpotential in the F-theory and the Type II/F theory duality which connects them leading a way to study the complicate D-brane system. In subsection 2.2, we construct the Coulomb phase and the Higgs phase of the D-brane system in terms of toric geometry which correspond to the 4-folds for F-theory compactification. In subsection 2.3, a brief review of the generalized GKZ system and the local solutions to it has been given. These solutions closely relate to the mirror map and the potential functions of the D-brane system in the paper.
In section 3, we apply the formulism to the parallel phase and the coincident phase of three D-brane systems. The superpotentials for the D-brane geometries are obtained near the large complex structure point in terms of open-closed deforming space. The Ooguri-Vafa invariants in parallel and coincident phases are extracted and first several orders of them are displayed with figures for comparison. Some observations are discussed.
The last section is a brief summary and further discussions.
\section{Superpotentials, toric geometry and differential equations}
\subsection{Superpotantial of brane and flux in type-II and F-theory}
Considering the space-filling D5-branes wrap on reducible curve $C=\sum_i\mathcal{C}_i$ embedded in a divisor $\mathcal{D}$ of Calabi-Yau 3-fold $M_3$, the effective superpotential is captured by the relative period which is defined by integration of the holomorphic $(3,0)$ form $\Omega^{(3,0)}(z,\hat{z})$ on an element $\gamma$ in relative homology group $H_3(M_3,\mathcal{D})$:
\be
\mathcal{W}_{\mathcal{N}=1}(z,\hat{z})=\Pi_\gamma(z,\hat{z})=\int_\gamma\Omega^{(3,0)}(z,\hat{z}),~~~\gamma\in H_3(M_3,\mathcal{D}).
\ee
The notations $z$ and $\hat{z}$ are for closed and open moduli respectively.
Thus the four-dimensional effective superpotential has an unified expression as a general linear combination of the integral of the basis of relative period \cite{LMW.02.1,LMW.02.2}:
\be\label{wob}
\mathcal{W}_{\mathcal{N}=1}(z,\hat{z})=\sum N_\alpha\Pi_\alpha(z,\hat{z})=\mathcal{W}_{open}(z,\hat{z})+\mathcal{W}_{closed}(z),
\ee
where the coefficients $N_\alpha$ are determined by the topological charges of branes and background flux, and $\Pi_\alpha(\zz)$ are the relative period integrals in both of open and closed sectors.
The superpotential can be split into two sectors, namely the open sector contributed purely by the open strings and the closed sector contributed by the 3-form flux superpotential $W_{flux}=\int_{M_3}G\wedge \Omega$ of background flux $G$ \cite{GVW.99,TV.00}.
These two sectors have following form:
\be
\mathcal{W}_{open}(\zz)=\sum_{\gamma_\alpha,\partial\gamma_\alpha\neq0}\hat{n}_\alpha\hat{\Pi}_\alpha(\zz),~~~~~~~~~\mathcal{W}_{closed}(z)=\sum_{\gamma_\alpha,\partial\gamma_\alpha=0}n_\alpha\Pi_\alpha(z),
\ee
where $\gamma_\alpha$ is a set of basis for $H_3(M_3,\mathcal{D})$.
Above superpotential is called off-shell superpotential, since it depends on the open moduli.
The on-shell superpotentials can be defined from the normal function point of view \cite{W.06,MW.07} which is a special case when the D-branes wrapping on holomorphic curves leading to the critical value $d\mathcal{W}=0$ of off-shell superpotentials with respect to the open-string moduli $\hat{z}$. In this paper we focus on the off-shell superpotential.

While the mirror symmetry connects the the $N=1$ superpotential $\mathcal{W}$ of B-branes wrapped on complex submanifold of a Calabi-Yau threefold $M_3$ with the superpotential of the A-branes wrapped on the special Lagrangian submanifold of $W_3$ (the mirror of $M_3$), the superpotential on the B model side is related to the flux superpotential of a F-theory compactification $M_4$ (a dual fourfold of $M_3$).
The superpotentials of 4-form flux $G_4$ in F-theory compactified on the Calabi-Yau 4-fold $M_4$ is a section of the line bundle over the complex structure moduli space $\mathcal{M}_{CS}(M_4)$.
It enjoys the name Gukov-Vafa-Witten superpotential \cite{GVW.99} which has general form \cite{AHJMMS.09,JMW.09}
\be\label{wof}
\mathcal{W}_{GVW}(M_4)=\int_{M_4}G_4\wedge\Omega^{(4,0)}=\sum_\Sigma N_\Sigma(G_4)\Pi_\Sigma(\zz)+\mathcal{O}(g_s)+\mathcal{O}(e^{-1/g_s}),
\ee
where the $g_s$ is the string coupling strength and the leading term on the right hand side is the D-brane superpotential $\mathcal{W}_{N=1}$ (\ref{wob}).
Note that the linear combination coefficients of relative periods depend on the background flux $G_4$ on $M_4$.

Before we construct the $M_4$, let's start with the observation of a circle of dual relations presented below:
\be\label{msd1}
\xymatrix{
{\buildrel {\displaystyle(W_3,L)}\over {\rm (A-branes)}\hskip 20pt }\ar[d]_{4f}^{dual} \ar[r]^{mirror}_{symmetry}
&
\ar[l]\hskip 20pt{\buildrel {\displaystyle(M_3,\mathcal{D})}\over {\rm (B-branes)}}\ar[d]_{4f}^{dual} \cr
\hskip 25pt \tilde{W}_4 \hskip 20pt\ar[r]^{mirror}_{symmetry}&\hskip 20pt \ar[l]\tilde{M}_4}
\ee
The dual relation on the right hand side of this diagram identifies the relative periods of the brane geometry $(M_3,\mathcal{D})$ with the periods of holomorphic $(4,0)$ form on the non-compact 4-fold $\tilde{M}_4$,
as they are governed by the same Picard-Fuchs equations which we would introduce in the subsection 2.3.
It is the consequence of the isomorphism between the moduli space $\mathcal{M}(M_3,\mathcal{D})$ of open-closed system and the moduli space $\mathcal{M}(\tilde{M}_4)$ of the non-compact 4-fold $\tilde{M}_4$ which depends on the `closed' moduli only \cite{AHJMMS.09}.
The following diagram shows the structure of mirror pair $(\tilde{W}_4,\tilde{M}_4)$.
\be\label{somof}
\xymatrix{
  W_3 \ar[r]_{} &\hskip 10pt \tilde{W}_4 \ar[d]_{\pi} \hskip 10pt \ar[r]^{mirror}_{symmetry}  &\hskip 10pt \tilde{M}_4 \hskip 10pt \ar[d]_{\pi} \ar[l] & \tilde{\mathcal{E}} \ar[l] \\
    &{{T}}  & M_3 &   }
\ee
Generally, the geometry of $\tilde{M}_4$ is a elliptic fibration over the 3-fold $M_3$ \cite{BVS.95,KMP.96,AKLM.99},
while it's mirror partner $\tilde{W}_4$ is a fibration basing on a disk $T$ with fiber a Calabi-Yau 3-fold of type $W_3$.
However the mirror pair of fourfolds $(\tilde{W}_4, \tilde{M}_4)$ are non-compact, they have to be compactified to a pair of compact 4-folds $(W_4,M_4)$ for the honest 4 dimensional F-theory compactification. The compact 4-fold $W_4$ is constructed by a simple ${P}^1$ compactification of the non-compact base $T$ of $\tilde{W}_4$, while the compact 4-fold $M_4$ is the mirror of $W_4$ and it is dual to the B-brane geometry $(M_3, \mathcal{D})$.
\be\label{DMS}
\xymatrix{
{W_4\hskip 10pt } \ar[d]_{Vol(P^1)\rightarrow\infty}\ar[r]^{mirror}_{symmetry}
&\hskip 10pt M_4\ar[l]\ar[d]^{g_s\rightarrow 0}  \cr
 \tilde{W}_4 \hskip 10pt\ar[r]^{mirror}_{symmetry}&\hskip 10pt \tilde{M}_4\ar[l]}
\ee

It is important to identify the image of the large base limit of $W_4$ in moduli space of the F-theory compactification on the $M_4$.
Presenting in the diagram (\ref{DMS}), when volume of $P^1$ go to the infinity, the corresponding behaviour on the B-model side is the weak coupling limit as the $g_s$ go to zero. This limit bring the mirror pair back to the non-compact states, meanwhile give us the D-brane superpotential $\mathcal{W}_{N=1}$ from the GVW superpotential $W_{GVW}$ of F-theory as follows:
\be\label{lim_of_sc}
\mathcal{W}_{\mathcal{N}=1}(M_3,\mathcal{D})=\sum N_\Sigma(G_4)\Pi_\Sigma(\zz)=\lim_{g_s\rightarrow 0}\mathcal{W}_{GVW}(M_4).
\ee
The GVW superpotential computes the D-brane superpotentials at the strict limit $g_s=0$, where most of the degree of freedom decouple from the superpotential sector.
The weak coupling limit in fact is a constrict on the moduli space $\mathcal{M}(M_4)$ inducing to the subspace of it which identifies with $\mathcal{M}(\tilde{M}_4)$, isomorphism of $\mathcal{M}(M_3,\mathcal{D})$, while on the A-side the corresponding point in the moduli space $\mathcal{M}(W_4)$ is the large base limit $Vol(\mathbf{P}^1)\rightarrow \infty$ \cite{BM.98,BM.05}.
Although the high order of $g_s$ terms are switched off at the decoupling point, the information of $\mathcal{N}=1$ superpotential remains in the leading term.
On the A-model side of mirror pair $(W_4,M_4)$, the K\"{a}hler volume of the topological cycles $\pi_4$ in $H_4(W_4,{Z})$ computes the 'classical terms' of the flux superpotential, which match the double logarithmic terms of flux $(G_4)$ superpotential on the B-model side as the solutions of generalized GKZ system\cite{AHJMMS.09}.
The matching of classical terms determines the bulk potential and the superpotential among the periods integrals on the B-model side.

\subsection{Enhanced polyhedra for Coulomb/Higgs phase}

The Calabi-Yau manifolds we discuss in this paper are defined by the hypersurfaces in ambient space which is toric variety. We refer to \cite{HLY.96,HKTY.95.1,HKTY.95.2,B.93} for more detail about the background of toric geometry. The notation is as follows: $(\nabla_4,\Delta_4)$ is a pair of mutually reflective polyhedrons and $(W_3,M_3)$ is a mirror pair of Calabi-Yau 3-fold defined by the hypersurfaces in ambient toric varieties $(P_{\Sigma(\nabla_4)},P_{\Sigma(\Delta_4)})$.
The $P_{\Sigma(\Delta_4)}$ is the toric variety with fan $\Sigma(\Delta_4)$ defined by a set of cones over the faces of $\Delta_4$. The hypersurfaces $M_3$ is defined by $p$ integral points of $\nabla_4$ as the zero locus of the polynomial $P$ in the $P_{\Sigma(\Delta_4)}$
\be\label{pop}
P=\sum_{i=0}^{p-1}a_i\prod_{k=1}^4X_k^{v^*_{i,k}},
\ee
where $v^*_{i,k}$ means the $k$-th coordinate of the integral point $v^*_i$ in $\nabla_4$ and the $X_k$ are local coordinates on a open torus $({C}^*)^4\subset P_{\Sigma(\Delta_4)}$. The coefficients $a_i$ are complex parameters related to the complex structure of $M_3$. In the homogeneous coordinates $x_j$ on the toric ambient space, $P$ can be rewritten as
\be
P=\sum_{i=0}^{p-1}a_i\prod_{v\in\Delta_4}x_j^{\langle v,v_i^*\rangle+1}.
\ee

The $n$ parallel D-branes are defined by reducible divisor:
\be
\begin{split}
Q(\mathcal{D})=&\prod_{m=0}^n(\phi_ma_0+a_i\prod_{k=1}^4X_k^{v^*_{i,k}})\\
=&\sum_{k=0}^nb_k\prod_{v\in\Delta_4}x_j^{k\langle v,v^*_i\rangle+n},
\end{split}
\ee
whose irreducible components lie in a single parameter family of divisor $\mathcal{D}_s\equiv\phi a_0+a_i\prod_{k=1}^4X_k^{v^*_{i,k}}$.
The parameters $b_k$ encode the open-string deformations of the $n$ parallel branes. In other words, the brane component deformation parameters $\phi_m$ describes the position of the $m$th individual D-brane component.
The parallel D-brane geometry corresponds to the Coulomb phase of the gauge theory which gives rise to the $U(1)\times U(1) \times ...\times U(1)$ group, product of $n$ $U(1)$'s, and each $U(1)$ group describes electromagnetism including the Coulomb field.
Similar to the construction in ref.~\cite{AHMM.09}, the combinatoric data of the parallel D-branes phase can be recovered in an one dimensional higher polyhedron $\tilde{\nabla}_5$ defining the non-compact 4-fold $\tilde{W}_4$. Then the relevant vertices shaping the $\tilde{\nabla}_5$ are

\be\label{popb}
\tilde{v}_j^*=\begin{cases}(v_j^*,0)& j=0,...,p-1,\\
(mv_i^*,1)& j=p+m, 0\leq m\leq n.\end{cases}
\ee

When parallel D-branes coincide and dissolved into each other, the $U(1)\times U(1) \times ...\times U(1)$ gauge group obtain an enhancement becoming $SU(n)$ group corresponding to a phase transition to the Higgs branch of the gauge theory.
Geometrically, the non-Abelian gauge group relates to the singularities on the Calabi-Yau manifolds.
In Toric language the singular curve corresponds to a one-dimensional edge of the dual polyhedron with integral lattice points on it. The resolution process is standard in terms of toric geometry \cite{F.Introduction}. It adds all the interior points on the one-dimensional edges into the points configuration and each of these vertices corresponds an exceptional divisor in the blow-up of the Calabi-Yau manifolds.
Notice that in the Coulomb phase we add $n+1$ new points lying on the one-dimensional edge for the $n$ parallel D-branes and the $n-1$ interior points on the edge rightly resolve the $Z_n$ singularity.
Furthermore, for the $Z_n$ curve singularity the intersection matrix of the exceptional divisors is proportional to the Cartan matrix of $A_{n-1}$ with self-intersections normalized to $-1$.
Inversely, if we remove the $n-1$ interior points, we could recover the singular 4-fold which gives rise to the enhanced gauge symmetry $SU(n)$ matching the expectation in the coincident D-brane system \cite{BV.95}.

In addition to compactify the non-compact fourfold $\tilde{W}_4$, we add one more point into the points configuration beneath the hyperplane $Y=\{v \in R^5|v_5=0\}$.
Together with the points for D-brane geometry, all the points define the enhanced polyhedron $\nabla_5$ corresponding to the compact Calabi-Yau 4-fold $W_4$. The details are presented in the section 3.

\subsection{Generalized GKZ systems, relative periods and local solutions}

The relative periods satisfy a set of differential equations named Picard-Fuchs equations which share the solutions with a larger system, namely the generalized hypergeometric GKZ system. It has the following form:
\be
\mathcal{L}(l^a)(\vartheta,\hat{\vartheta})\Pi(\zz)=0,
\ee
where $\vartheta$ and $\hat{\vartheta}$ are abbreviations of $z\partial_{z}$ and $\hat{z}\partial_{\hat{z}}$.
And the differential operators of the Picard-Fuchs equations can be derived from the GKZ system as follows:
\be
\mathcal{L}(l^a)=\prod_{k=1}^{l^a_0}(\vartheta_0-k)\prod_{l^a_j>0}\prod_{k=0}^{l^a_j-1}(\vartheta_j-k)-(-1)^{l^a_0}z_a\prod_{k=1}^{-l^a_0}(\vartheta_0-k)\prod_{l^a_j<0}\prod_{k=0}^{-l^a_j-1}(\vartheta_j-k),
\ee
where $\vartheta_j=a_j\frac{\partial}{\partial a_j}$ are logarithmic derivatives with respect to the parameters $a_j$,
and the $l^a$ are the generators of Mori cone \cite{FS.03,F.01,S.07,R.07} of $P_{\Sigma(\nabla_5)}$.
They are also been known as the charge vectors of the gauged linear sigma model (GLSM)\cite{Wi.93} for $a=1,...,k=h^{1,1}(W_4)$.
Remarkably, on one hand the charge vectors $l^a$ correspond to the maximal triangulation of $\nabla_5$. At the vicinity of the large complex structure limit point in the complex structure moduli space of $M_4$, the local coordinates are given by
\be\label{ac}
z_a=(-1)^{l_0^a}\prod_ja_j^{l_j^a}.
\ee
They are torus invariant algebraic coordinates on the large complex structure phase, where the non-perturbative instanton corrections are suppressed exponentially in.
On the other hand once the generators of Mori cone have been chosen, the generators of K\"{a}hler cone are also determined result from the duality of Mori cone and K\"{a}hler cone. In other word, the basis $J_a$ of $H^{1,1}(W_4)$ can be settled down dual to the $l^a$ and naturally give the local coordinates $t_a$ around the large radius limit point mirror to the large complex structure point. The coordinates $t_a$ are also known as the flat coordinates.

On the F-theory side, the GKZ system is derived from the combinatoric data of the 5 dimensional polyhedra corresponding to the fourfolds. So the solutions relates to the periods of the fourfolds which equalize the open and closed string moduli.
On the Type II theory side, the enhanced polyhedra corresponding to the D-brane geometry give rise to the GKZ system whose solutions encode the relative periods giving rise to the mirror maps and the physical potentials.

In terms of the periods of the fourfolds, according to the ref.~\cite{HLY.96}, the local solutions to the GKZ system can be derived from the fundamental period
\be\label{fw}
w_0(z;\rho)=\sum_{m_1,...,m_k\geq 0}\frac{\Gamma(-\sum (m_k+\rho_k)l_0^a+1)}{\prod_{1\leq i\leq p}\Gamma (\sum(m_k+\rho_k)l_i^a+1)}z^{m+\rho},
\ee
using the Frobenius method. The whole periods vector reads
\be\label{pmob}
\vec{\Pi}(z)=\begin{pmatrix} \Pi_0=&w_0(z;\rho)|_{\rho=0} \\
\Pi_{1,i}=&\partial_{\rho_i}w_0(z;\rho)|_{\rho=0}\\
\Pi_{2,n}=&\sum_{i_1,i_2}K_{i_1i_2;n}\partial_{\rho_{i_1}}\partial_{\rho_{i_2}}w_0(z;\rho)|_{\rho=0}\\
&\vdots\end{pmatrix},
\ee
where $n\in \{1,...,h\}$ and $h$ stands for the dimension of $H^4(\tilde{M}_4)$. The $K_{i_1i_2;n}$ are the combinatoric coefficients of the second derivative of $w_0$.
Guided by the mirror hypothesis, the period vector $\vec{\Pi}(z)$ has a dual description on the A-side with the general form
\be\label{pmoa}
\vec{\Pi}^*(t)=\begin{pmatrix} \Pi^*_0=&1\\
\Pi^*_{1,i}=&t_i\\
\Pi^*_{2,n}=&\sum_{i_1,i_2}K^*_{i_1i_2;n}t_{i_1}t_{i_2}+b_n+F^{inst}_n\\
&\vdots \end{pmatrix},
\ee
where $t_i=\Pi_{1,i}/\Pi_0$ are the flat coordinates. The coefficients $K^*_{i_1i_2;n}$ of leading terms relating to the classical sector in the periods can be easily determined once the 4-cycle $\pi^*_4 \in H_4(W_4,{Z})$ relating to the 4-form flux $G_4$ has been selected. The constants $b_n$ are not relevant in our discussion and the $F^{inst}_n$ stand for the instanton correction sector of the solutions. We demonstrate the detail in the section 3.

In terms of the relative periods of D-brane geometry, the entries of the relative period vector give rise to the mirror map, bulk potential and superpotential. The relative period vector is of the form
\be\label{pv}
\vec{\Pi}^*(t,\hat{t})=(1,t,\hat{t},F_{t}(t),\mathcal{W}(t,\hat{t})...)),
\ee
where $t(\hat{t})$ are the flat coordinates on the moduli space. The solution $F_{t}(t)\equiv\partial_{t}F(t)$ is determined only by the closed moduli, where $F(t)$ is the $N=2$ prepotential. It's corresponding part in the open sector is the superpotential $\mathcal{W}(t,\hat{t})$.
The flat coordinates could be found at large radius regime of A-model and the large complex structure regime of B-model and it defines the mirror map as follows
\be
t_i(z)=\frac{\Pi_{1,i}(z)}{\Pi_0},~~~\hat{t}_i(\zz)=\frac{\Pi_{1,i}(\zz)}{\Pi_0},
\ee
where $\Pi_0$, $\Pi_{1,i}(z)$ and $\Pi_{1,i}(\zz)$ are selected basis of the relative periods with respect to the open-closed moduli.
The instanton corrections are encoded as a power series expansion of $q_i=exp(2\pi i t_i)$ and $\hat{q}_i=exp(2\pi i\hat{t}_i)$:
\be\label{instex}
F^{inst}(t,\hat{t})=\sum_{\vec{k},\vec{m}}G_{\vec{k},\vec{m}}q^{\vec{k}}\hat{q}^{\vec{m}}=\sum_n\sum_{\vec{k},\vec{m}}\frac{N_{\vec{k},\vec{m}}}{n^2}q^{n\vec{k}}\hat{q}^{n\vec{m}}.
\ee
In the Eq.(\ref{instex}), $\{G_{\vec{k},\vec{m}}\}$ are open Gromov-Witten invariants labeled by relative homology class, where $\vec{m}$ represent the elements of $H_1(L)$ and $\vec{k}$ represent the elements of $H_2(W_3)$, and $\{N_{\vec{k},\vec{m}}\}$ are Ooguri-Vafa invariants.

\section{Models}
In this section, we study the parallel phase and the coincident phase, namely the Coulomb and Higgs branch respectively, in three different D-brane models.
Previously, it should be point out that we select the same compactifying point $\tilde{v}^*_c=(0,0,0,0,-1)$ in terms of combinatoric data in all the cases\footnote{The compactfying point is different from the choice in the \cite{AHJMMS.09} which picked other integral point on the hyperplane $Y=\{v\in {R}^5|v_5=-1\}$. However, the detail of the ${P}^1$ compactification dominates the subleading terms in $g_s$ and would be irrelevant in the decoupling limit.}.
\subsection{D-branes on the mirror quintic}
The mirror of Quintic is defined as a degree 5 hypersurface $P$ in the ambient toric variety $P_{\Sigma(\Delta_4)}$ determined by the vertices of the polyhedron $\Delta_4$
\be
\begin{split}
&v_1=(4,-1,-1,-1), v_2=(-1,4,-1,-1), v_3=(-1,-1,4,-1),\\
&v_4=(-1,-1,-1,4),v_5=(-1,-1,-1,-1).
\end{split}
\ee
The integral points in dual polyhedra $\nabla_4$ are
\be
\begin{split}
&v^*_0=(0,0,0,0), v^*_1=(1,0,0,0), v^*_2=(0,1,0,0), v^*_3=(0,0,1,0),\\
&v^*_4=(0,0,0,1), v^*_5=(-1,-1,-1,-1),
\end{split}
\ee
leading to the defining polynomial for the toric hypersurface
\be
P=a_1x_1^5+a_2x_2^5+a_3x_3^5+a_4x_4^5+a_5x_5^5+a_0x_1x_2x_3x_4x_5.
\ee

\subsubsection{Parallel D-branes phase}
We consider the parallel branes which are described by the reducible divisor $\mathcal{D}=\mathcal{D}_1+\mathcal{D}_2$ realized by the degree 8 homogeneous equation
\begin{alignat}{2}\label{pl}
Q&=b_0(x_1x_2x_3x_4x_5)^2+b_1x_1^6x_2x_3x_4x_5+b_2x_1^{10}\\
 &\sim\prod_{i=1}^2(\phi_ia_0x_1x_2x_3x_4x_5+a_1x_1^5).
\end{alignat}
By factorizing the polynomial (\ref{pl}),
the individual D-brane parameters $\phi_i$ display the ${Z}_2$ symmetry for two parallel branes corresponding to the two factors in the (3.5).
According to the construction in subsection 2.2, the open-closed system is encoded in the enhanced polyhedron $\tilde{\nabla}_5$ whose vertices are
\be
\begin{split}{\label{V1}}
&\tilde{v}_0^*=(0,0,0,0,0), \tilde{v}_1^*=(1,0,0,0,0), \tilde{v}_2^*=(0,1,0,0,0),\\
&\tilde{v}_3^*=(0,0,1,0,0), \tilde{v}_4^*=(0,0,0,1,0), \tilde{v}_5^*=(-1,-1,-1,-1,0),\\
&\tilde{v}_{6}^*=(0,0,0,0,1),\tilde{v}^*_7=(1,0,0,0,1),\tilde{v}^*_8=(2,0,0,0,1).
\end{split}
\ee
The geometry for the F-theory compactification, the compact 4-fold $W_4$, is defined by the polyhedron $\nabla_5$ with vertices given in (\ref{V1}) and $\tilde{v}^*_c$.

The generators of Mori cone of the toric variety determined by $\nabla_5$ are given by:
\begin{equation}{\label{L1}}
\begin{array}{ccccccccccccccccccccccccc}
     & &  & 0 &1&2&3&4&5&6&7&8&c & \\
  l^1&=&( & -4&0&1&1&1&1&-1&1&0&0 &)\\
  l^2&=&( & 0 &0&0&0&0&0&1&-2&1&0 &)\\
  l^3&=&( &-1&1&0&0&0&0&0&1&-1&0 &)\\
  l^4&=&( &0&-2&0&0&0&0&0&0&1&1 &).
\end{array}
\end{equation}
The K\"{a}hler form is $J=\sum_ak_aJ_a$, where $J_a$ denotes the basis of $H^{1,1}(W_4)$ dual to the Mori cone generated by (\ref{L1}). And $k_a$ are flat coordinates on the K\"{a}hler moduli space of the mirror four-fold $W_4$.
Then we choose the basis elements of $H_4(W_4)$ which are defined by intersections of the toric divisors $D_i$ corresponding to the $\tilde{v}_i^*$, namely
\be
\gamma_1=D_1\cap D_2,~\gamma_2=D_2\cap D_c,~\gamma_3=D_2\cap D_6,~\gamma_4=D_7\cap D_8.
\ee
After changing the variables as follows to visualize the closed and open moduli
\be
t=k_1+k_2+k_3,~~\hat{t}_1=k_2+k_3,~~\hat{t}_2=k_3,
\ee
the leading terms of the periods are
\be\label{leading_term1}
\tilde{\Pi}^*_{2,1}=5tk_4,~~\tilde{\Pi}^*_{2,2}=\frac{5}{2}t^2,~~\tilde{\Pi}^*_{2,3}=2(t-\hat{t}_1)^2,~~\tilde{\Pi}^*_{2,4}=2(t-\hat{t}_2)^2.
\ee
The classical period $\tilde{\Pi}^*_{2,1}$ depends on $k_4$ which is a K\"{a}hler parameter come from the K\"{a}hler volume of the base ${P}^1$ relating to the strength of string coupling. It wouldn't contribute to the potential functions when $Im~k_4$ goes to infinity in weak coupling limit.
The $\tilde{\Pi}^*_{2,2}$ only depends on the close moduli $t$ and is supposed to be the leading term of the bulk potential function $F_{t}(t)$, while $\tilde{\Pi}^*_{2,3},~\tilde{\Pi}^*_{2,4}$ are supposed to lead the D-brane superpotential $\mathcal{W}(t,\hat{t})$ which depends on open $(\hat{t})$ and closed $(t)$ parameters both.
It is more interesting that $\tilde{\Pi}^*_{2,3},~\tilde{\Pi}^*_{2,4}$ shows ${Z}_2$ symmetry with respect to the open moduli $\hat{t}_1$ and $\hat{t}_2$ matching the expectation that the superpotentials contributed by the two ${Z}_2$ symmetric parallel branes should inherit the symmetry.

The instanton corrections of the period integrals are recorded in the solution of the generalized GKZ system corresponding to the enhanced polyhedron $\tilde{\nabla}_5$. We identify the bulk potential and the superpotentials with the exact solutions to the GKZ system which lead by $\tilde{\Pi}^*_{2,2}, \tilde{\Pi}^*_{2,3}$ and $\tilde{\Pi}^*_{2,4}$ respectively.
Following the Eq.(\ref{fw}), in terms of algebraic coordinates (\ref{ac})
\be
z_1=\frac{a_2a_3a_4a_5b_1}{a_0^4b_0},~~~z_2=\frac{b_0b_2}{b_1^2},~~~z_3=\frac{a_1b_1}{a_0b_2},
\ee
the fundamental period and the logarithmic periods,
\begin{equation}
\Pi_0(z)=w_0(z;0),~\Pi_{1,i}(z)=\partial_{\rho_i}w_0(z;\rho)|_{\rho_i=0},~\Pi_{2,n}(z)=\sum_{i,j}K_{i,j;n}\partial_{\rho_i}\partial_{\rho_j}w_0(z;\rho)|_{\rho=0},
\end{equation}
solve the generalized GKZ system governed by charge vectors (\ref{L1}).
The flat coordinates are given by
\be
k_i=\frac{\Pi_{1,i}(z)}{\Pi_0(z)}=\frac{1}{2\pi i}log~z_i+...~.
\ee
Then the mixed inverse mirror maps in terms of $q_i=exp(2\pi ik_i)$ for $\{i=1,2,3\}$ are
\be
\begin{split}
z_1=&{q_1} + 24q_1^2 + {q_1}{q_2} + 24q_1^2q_2^2 - 2{q_1}{q_2}{q_3} +...\\
z_2=&{q_2} - 24{q_1}{q_2} + 972q_1^2{q_2} - 2q_2^2 + 120{q_1}q_2^2 - 5616q_1^2q_2^2 - {q_2}{q_3} + 24{q_1}{q_2}{q_3} - 972q_1^2{q_2}{q_3} + \\ &5q_2^2{q_3} - 240{q_1}q_2^2{q_3} - 3q_2^2q_3^2 +...\\
z_3=&{q_3} + {q_2}{q_3} - 48{q_1}{q_2}{q_3} + 972q_1^2{q_2}{q_3} + q_3^2 - 178{q_1}{q_2}q_3^2 + q_2^2q_3^2 +...~.
\end{split}
\ee
According to the leading terms (\ref{leading_term1}), we find the relative periods which corresponds to the closed-string period and D-brane superpotentials in the A-model as follows:

\be\label{spp1}
\begin{split}
F_{t}(t)\equiv\Pi_{2,2}=&\frac{5}{2}t^2+\frac{1}{4\pi^2}(2875q+\frac{4876875}{4}q^2+\frac{8564575000}{9}q^3+...)\\
\mathcal{W}_1(t,\hat{t}_1)\equiv\Pi_{2,3}=&2(t-\hat{t}_1)^2+\frac{2}{4\pi^2}( 800q + 340000q^2 + \frac{{2388816800}}{9}q^3 -160q\hat{q}_1^{-1} + \\
&6600q^2\hat{q}_1^{-2} + 10\hat{q}_1  - 58280q^2\hat{q}_1^{-1} +\frac{{\hat{q}_1^2}}{2}+ 1020q\hat{q}_1 -42557680q^3\hat{q}_1^{-1} +\\
&\frac{{\hat{q}_1^3}}{9} - 730q\hat{q}_1^2 + 532090q^2\hat{q}_1 +...)\\
\mathcal{W}_2(t,\hat{t}_2)\equiv\Pi_{2,4}=&2(t-\hat{t}_2)^2+\frac{2}{4\pi^2}( 800q+340000q^2+ \frac{{2388816800}}{9}q^3 - 160q\hat{q}_2^{-1} + \\
&6600q^2\hat{q}_2^{-2} + 10\hat{q}_2  - 58280q^2\hat{q}_2^{-1} + \frac{{\hat{q}_2^2}}{2}+ 1020q\hat{q}_2 -42557680q^3\hat{q}_2^{-1} + \\
&\frac{{\hat{q}_2^3}}{9} - 730q\hat{q}_2^2 + 532090q^2\hat{q}_2 +...),
\end{split}
\ee
where $q=exp(2\pi it)$, $\hat{q}_1=exp(2\pi i\hat{t}_1)$ and $\hat{q}_2=exp(2\pi i\hat{t}_2)$. Absolutely, the bulk potential $F_{t}(t)$ only depends on the closed modulus $t$ and the D-brane superpotentials $\mathcal{W}_1(t,\hat{t}_2), \mathcal{W}_2(t,\hat{t}_1)$ have kept the $Z_2$ symmetry with respect to $\hat{t}_1$ and $\hat{t}_2$ after adding the instanton correction meeting our anticipation.

To compare, here we also list the superpotential for the D-brane with one open deformation modulus \cite{AHJMMS.09} described by the divisor $\mathcal{D}_0=\phi_0a_0x_1x_2x_3x_4x_5+a_1x_1^5$ as follows:
\be\label{w01}
\begin{split}
\mathcal{W}_0(t,\hat{t}_0)=&2(t-\hat{t}_0)^2+\frac{2}{4\pi^2}( 800q+340000q^2+ \frac{{2388816800}}{9}q^3 - 160q\hat{q}_0^{-1} + \\
&6600q^2\hat{q}_0^{-2} + 10\hat{q}_0  - 58280q^2\hat{q}_0^{-1} + \frac{{\hat{q}_0^2}}{2}+ 1020q\hat{q}_0 -42557680q^3\hat{q}_0^{-1} + \\
&\frac{{\hat{q}_0^3}}{9} - 730q\hat{q}_0^2 + 532090q^2\hat{q}_0 +...),
\end{split}
\ee
where $\hat{t}_0$ is the only open modulus and $\hat{q}_0=exp(2\pi i\hat{t}_0)$.
It is noticeable that the $\mathcal{W}_1(t, \hat{t}_1)$ and $\mathcal{W}_2(t, \hat{t}_2)$ (\ref{spp1}) are identified with the superpotential $\mathcal{W}_0(t, \hat{t}_0)$ (\ref{w01}) when $\hat{t}_0=\hat{t}_1=\hat{t}_2$, namely
\be\label{rlt1}
\mathcal{W}_{1}(t,\hat{t}_{1})=\mathcal{W}_{2}(t,\hat{t}_{2})=\mathcal{W}_0(t,\hat{t}_0).
\ee

\subsubsection{Coincident D-branes phase}
According to the enhanced polyhedron $\nabla_5$, the defining polynomial of the the dual 4-fold $M_4$ on the B-model side is
\be
\begin{split}
\tilde{P}=&a_1x_1^5x_4x_5x_6x_7x_8x_9x_{10}^5+a_2x_2^5x_9^4+a_3x_3^5x_8^4+a_4x_4^5x_7^4+a_5x_5^5x_6^4+a_6x_1^2x_2^2x_3^2x_4^2x_5^2\\
&a_7x_1^6x_2x_3x_4x_5x_{10}^4+a_8x_1^{10}x_{10}^8+a_9x_6^2x_7^2x_8^2x_9^2x_{10}^2+a_0x_1x_2x_3x_4x_5x_6x_7x_8x_9x_{10}.
\end{split}
\ee
For simplifying the notation, we also denote the coefficients of the monomials in the polynomial as $a_i$'s, and the relation between the notations in the D-brane geometry and the 4-fold are as follows:
\be
a_i=\begin{cases}
a_i&0 \leq i\leq 5,\\
b_{i-6}&
6\leq i \leq 8,\\
c&
i=9.
\end{cases}
\ee
When $b_1^2=4b_0b_2$, the defining equation for the parallel D-branes becomes:
\be
Q\sim(\phi a_0x_2x_3x_4x_5+a_1x_1^4)^2,
\ee
which means the two individual D-branes coincide as $\phi_1=\phi_2=\phi$. Correspondingly, the equivalent description
\be\label{coincide1}
a_7^2=4a_6a_8
\ee
gives rise to the perfect square $(x_1x_2x_3x_4x_5\pm x_1^5x_{10}^4)^2$ in the $\tilde{P}$. Obviously $M_4$ becomes singular and the condition (\ref{coincide1}) constricts the complex moduli space of the 4-fold to a submanifold which is the coincident D-branes phase. The dual description of the coincident on the A-model side is the blow-down of the exceptional divisor developing the curve singularity on the $W_4$.

\begin{figure}[htbp]
  \centering
  \includegraphics[width=0.7\textwidth]{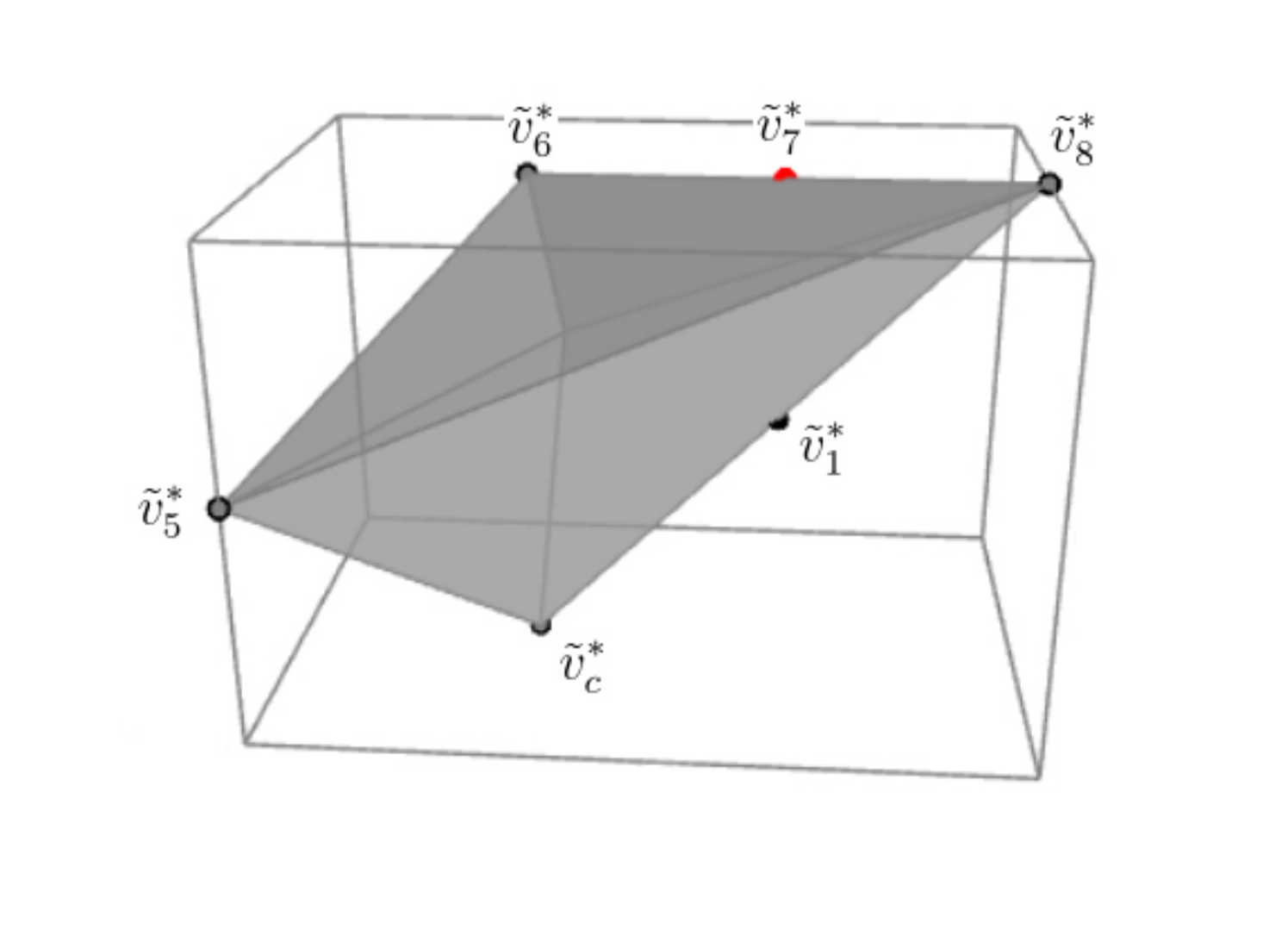}
  \caption{projected polyhedra of the $n=2$ coincident D-brane in the mirror quintic}\label{fig:1}
\end{figure}

The polyhedra of the $N=2$ coincident D-brane in the mirror quintic is projected on the hyperplane spanned by $x_1,x_2$ and $x_5$ in the Figure~\ref{fig:1}. The points $\tilde{v}^*_6, \tilde{v}^*_7, \tilde{v}^*_8$ lie on the one-dimensional edge and we ignore the interior point $\tilde{v}^*_7$ to recover the singularity corresponding to the coincidence of the branes.
It is noticeable that the interior points on the edge span the Dynkin diagram of the $A_1$ and the dual Calabi-Yau 4-fold develops the $A_1$ singularity when all the $2$ parallel D-branes coincide.

Then we obtain the new charge vectors (\ref{L2}) for the coincident D-branes phase corresponding to the maximal triangulation of the point configuration without $\tilde{v}^*_7$.
\begin{equation}\label{L2}
\begin{array}{ccccccccccccccccccccccccc}
     & &  & 0 &1&2&3&4&5&6&8&c& \\
  l^1&=&( & -8&0&2&2&2&2&-1&1&0&)\\
  l^2&=&( & -2 &2&0&0&0&0&1&-1&0&)\\
  l^3&=&( &0&-2&0&0&0&0&0&1&1&)
\end{array}
\end{equation}
The K\"{a}hler form is $J=\sum_ak_aJ_a$, where $J_a$ denotes the basis of $H^{1,1}(W_4)$ dual to the Mori cone generated by (\ref{L2}). And $k_a$ are flat coordinates on the K\"{a}hler moduli space of the mirror four-fold $W_4$.
Then we choose basis elements (\ref{basis312}) of $H_4(W_4)$ which are defined by intersections of the toric divisors $D_i$ corresponding to the $\tilde{v}_i^*$.
\be\label{basis312}
\gamma_1=D_1\cap D_2,~\gamma_2=D_2\cap D_c,~\gamma_3=D_2\cap D_6
\ee
After transforming the variables as follows
\be
t=k_1+k_2,~~\hat{t}=k_2,
\ee
the leading terms of the period integrals are
\be\label{leading_term2}
\tilde{\Pi}^*_{2,1}=5tk_3,~~\tilde{\Pi}^*_{2,2}=\frac{5}{2}t^2,~~\tilde{\Pi}^*_{2,3}=2(t-\hat{t})^2.
\ee
In (\ref{leading_term2}), the $\tilde{\Pi}^*_{2,2}$ depends on the closed modulus $t$ purely leading the bulk potential, and the $\tilde{\Pi}^*_{2,3}$ rely on closed ($t$) and open ($\hat{t}$) modulus both leading the superpotential. It is noticeable that there is only one open modulus in the coincident branes phase, since the coinciding condition reduces the degree of freedom of the open-closed parameter space. The open deformation $\hat{t}$ can be interpreted as the position parameter of the two coincident D-branes.

The instanton corrections of the period integrals can be recovered from the solution of the generalized GKZ system corresponding to the enhanced polyhedron $\tilde{\nabla}_5$.
Following the Eq.(\ref{fw}), in terms of algebraic coordinates (\ref{ac})
\be
z_1=\frac{a_2^2a_3^2a_4^2a_5^2a_8}{a_0^8a_6},~~~z_2=\frac{a_1^2a_6}{a_0^2a_8},
\ee
the fundamental period and the logarithmic periods,
\begin{equation}
\Pi_0(z)=w_0(z;0),~\Pi_{1,i}(z)=\partial_{\rho_i}w_0(z;\rho)|_{\rho_i=0},~\Pi_{2,n}(z)=\sum_{i,j}K_{i,j;n}\partial_{\rho_i}\partial_{\rho_j}w_0(z;\rho)|_{\rho=0},
\end{equation}
solve the generalized GKZ system governed by charge vectors (\ref{L2}).
The flat coordinates are given by
\be
k_i=\frac{\Pi_{1,i}(z)}{\Pi_0(z)}=\frac{1}{2\pi i}log~z_i+...~.
\ee
Then the mixed inverse mirror maps in terms of $q_i=exp(2\pi ik_i)$ for $\{i=1,2,3\}$ are
\be
\begin{split}
z_1=&{q_1} + 2520q_1^2 - {q_1}{q_2} - 1298880q_1^2{q_2} + 486450q_1^2q_2^2 + ...\\
z_2=&{q_2} - 2520{q_1}{q_2} + 28356300q_1^2{q_2} + q_2^2 - 326610{q_1}q_2^2 - 5279337000q_1^2q_2^2 +...~.
\end{split}
\ee
We identify the bulk potential and the superpotentials (\ref{potential312}) with the exact solutions to the GKZ system which lead by $\tilde{\Pi}^*_{2,2}$ and $\tilde{\Pi}^*_{2,3}$ respectively, where $q = exp(2\pi it)$.
\be\label{potential312}
\begin{split}
F_t(t)=&\frac{5}{2}t^2+\frac{1}{4\pi^2}(\frac{{7013625q}}{2} - \frac{{10326895984375{q^2}}}{{72}} + \frac{{9387844880634090775{q^3}}}{3} + ...)\\
\mathcal{W}_c(t,\hat{t})=&2(t-\hat{t})^2 +\frac{1}{4\pi^2}(- 14688{q_1} + \frac{{26{q_2}}}{3}  + \frac{{473866120q_1^2}}{3} + 1949616{q_1}{q_2} + \frac{{17q_2^2}}{{18}} -\\ &\frac{{18237918449856q_1^3}}{5}   - \frac{{98028296480}}{3}q_1^2{q_2}  + 2532460{q_1}q_2^2  - \frac{{47q_2^3}}{{135}} +...)
\end{split}
\ee

\subsubsection{The disk invariants of two phases}
In subsection 3.1.1, we find two independent solutions to the GKZ system corresponding to the superpotentials $\mathcal{W}_1(t, \hat{t}_1)$ and $\mathcal{W}_2(t, \hat{t}_2)$ for the two parallel D-branes which rely on the open moduli $\hat{t}_1$ and $\hat{t}_2$ respectively.
As a result of that the two parallel D-branes live in the same family of divisors, the superpotentials of them present the $Z_2$ symmetry in terms of exchanging $\hat{t}_1$ and $\hat{t}_2$ which locate the two parallel D-branes respectively and according to the Eq.(\ref{rlt1}) when $\hat{t}_0=\hat{t}_1=\hat{t}_2$ the $\mathcal{W}_1(t, \hat{t}_1)$, $\mathcal{W}_2(t, \hat{t}_2)$ (\ref{spp1}) are identified with the superpotential $\mathcal{W}_0(t, \hat{t}_0)$ (\ref{w01}) which contains the only D-brane parameterized by $\hat{t}_0$.
This fact also suggests that there is no interaction between two topological D-branes in the parallel phase of the D-branes system of our consideration.
Therefor the two individual D-branes give rise to the superpotentials independently as what happens in the the single D-brane system and give the same Ooguri-Vafa invariants which are called $U(1)$ Ooguri-Vafa invariants.
The Table~\ref{tab:1} shows the predictions of $U(1)$ Oogrui-Vafa invariants for $\mathcal{W}_1(t,\hat{t}_1)$ according to Eq(\ref{instex}), while the $U(1)$ Ooguri-Vafa invariants extracted from the superpotential $\mathcal{W}_2(t,\hat{t}_2)$ is the same as the invariants listed in Table~\ref{tab:1} up to a transformation $n_1\rightarrow n_1,~n_2\rightarrow n_2=n_1,~n_3\rightarrow n_3$.
When the two parallel D-branes coincide, phase transition appears and the singularities arise on the dual F-theory Calabi-Yau 4-fold since the blow-down of the exceptional divisors. In terms of gauge theory on the worldvolume of the D-branes, the gauge group is enhanced to the $SU(2)$, and the Ooguri-Vafa invariants extracted from the coincident superpotential $\mathcal{W}_c(t,\hat{t})$ are called the $SU(2)$ Ooguri-Vafa invariants which are presented in Table~\ref{tab:2}.
As we mentioned, the invariants in the Table~\ref{tab:1} are identified with the Ooguri-Vafa invariants of the D-brane system with one of the two parallel D-branes. Thus the results also demonstrates the difference between the single D-brane and the coincident two D-branes in terms of topological D-brane system, although they are the same submanifold from the viewpoint of the set theory.

\begin{figure}[H]
  \setlength{\abovecaptionskip}{-1cm}
  \setlength{\belowcaptionskip}{-0.5cm}
  \centering
  \includegraphics[width=1\textwidth]{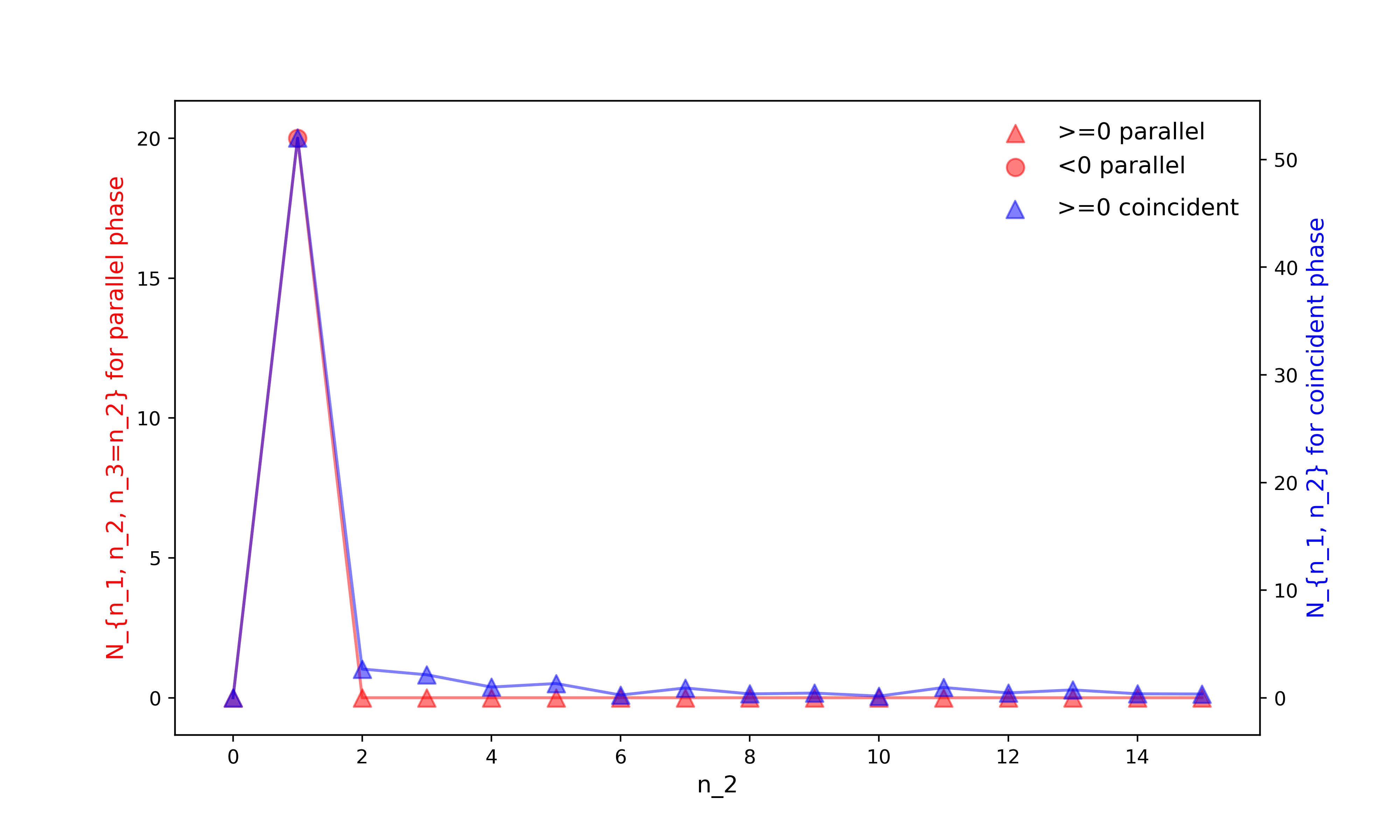}
  \caption{}\label{fig:2}
\end{figure}

\begin{figure}[H]
  \setlength{\abovecaptionskip}{-1cm}
  \setlength{\belowcaptionskip}{-0.5cm}
  \centering
  \includegraphics[width=1\textwidth]{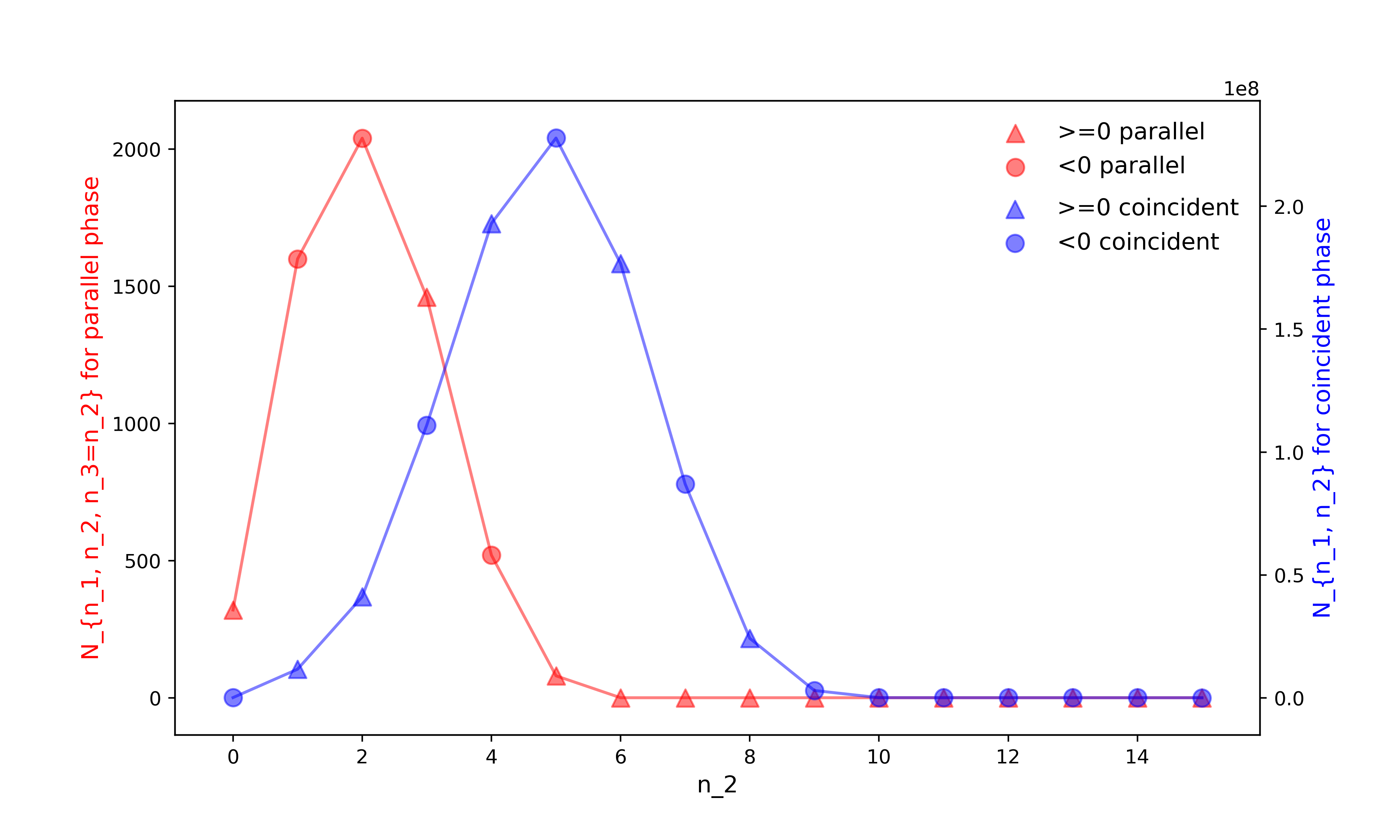}
  \caption{}\label{fig:3}
\end{figure}

\begin{figure}[H]
  \setlength{\abovecaptionskip}{-1cm}
  \setlength{\belowcaptionskip}{-0.5cm}
  \centering
  \includegraphics[width=1\textwidth]{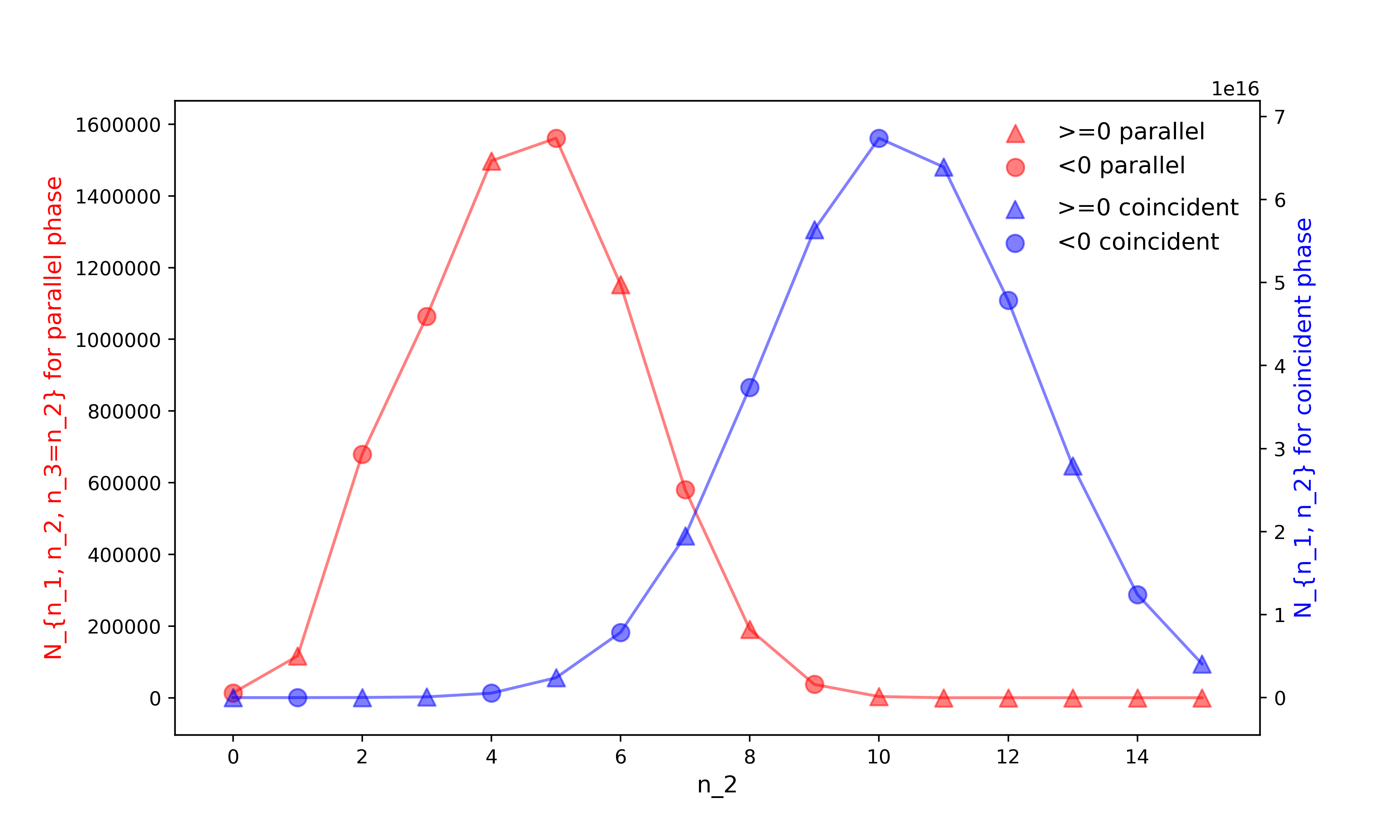}
  \caption{}\label{fig:4}
\end{figure}

For comparing the Ooguri-Vafa invariants of the two different phases, we display the first several terms of the absolute value of the $U(1)$ and $SU(2)$ Ooguri-Vafa invariants in one picture as the Figure~\ref{fig:2}, Figure~\ref{fig:3} and Figure~\ref{fig:4}. We mark the positive data with the Delta and the original negative data with the point, connecting them with red lines for the parallel phase and blue lines for the coincident phase. In addition, the different scales are used for the parallel and coincident phases on the left and right vertical axis respectively considering the huge difference of data for the two phase.  These markers are used in the figures appearing in all the following models.
In the three figures, the Ooguri-Vafa invariants for parallel phase and the coincident phase distribute like wave-packet both when the index $n_1$ has been fixed. Generally the value of these invariants climb to a peak and then decrease to zero (or near zero in the coincident case) with the increase of $n_2$.

In the Figure~\ref{fig:2} the index is fixed as $n_1=0$, the peak value of invariant for parallel phase is $N_{0,1,1}=20$ when $n_2=1$, while the peak value for coincident case is $N_{0,1}=54$ when $n_2=1$. The center of the two wave-packets are coincident and the duration of them are nearly the same. Since $n_2=2$ the red line goes along the horizontal axis $N_{n_1,n_2,n_3}=0$, and the blue line fluctuate around the red line slightly.

In the Figure~\ref{fig:3} as $n_1$ is fixed at 1, the peak values for parallel phase are $N_{1,2,2}=2040$ and for coincident phase  are $N_{1,5}=1594208432/7$ which is over 110000 times of $N_{1,2,2}$ . The center of the red wave-packet is at $n_2=2$ while the center of the blue one shift right to the $n_2=5$. The durations of the two wave-packets become to 6 (parallel phase) and 10 (coincident phase) respectively. Again, as the increase of $n_2$ the red line and blue line drop down near the position of zero and almost coincident.

In the Figure~\ref{fig:4} the indexes are fixed as $n_1=2$, when $n_2=n_3=5$ the value of invariant for parallel phase reach a peak $N_{2,5,5}=1561100$, the peak value for coincident case is $N_{2,10}=\frac{3844468233749481849712}{57057}$ which is nearly 40 billion times of the parallel phase peak. The red wave-packet (for parallel phase) is centered at $n_2=5$ with duration 10, while the blue one (for coincident phase) is centered at $n_2=10$ with duration 12.

To conclude, the value of peak for the coincident phase are bigger than the peak value for the parallel phase, and as the increase of the $n_1$ the difference of the peak values becomes extremely large. However, with the condition of various $n_1$, the $U(1)$ and $SU(2)$ invariants drop to near zero both as the free index $n_2$ increasing.
It should also be noticed that since $n_1=1$ the wave-packets for the two phases center at distinct locations. We can see when the two D-branes coincide the center of the wave-packets for the Ooguri-Vafa invariants seemly shift to the right side of the center for parallel phase, namely the direction of the larger $n_2$.
Moreover the coincident phase wave-packets have wider duration than ones for parallel phase.
From the point view of physics, the Ooguri-Vafa invariants are interpreted as the number of the BPS states. Our calculation demonstrate that when the parallel D-branes coincide the D-brane system generates totally different BPS states spectrum and in the coincident phase the structures of spectrum are more complicate. This phenomenon is a signal of the phase transition between the Coulomb branch and the Higgs branch.

\subsection{D-branes on hypersurface $X_8(1,1,2,2,2)$}
The degree 8 hypersurface in the ambient toric variety $P_{\Sigma(\Delta_4)}$ determined by the vertices of the polyhedron $\Delta_4$
\be
\begin{split}
&v_1=(7,-1,-1,-1), v_2=(-1,3,-1,-1), v_3=(-1,-1,3,-1),\\
&v_4=(-1,-1,-1,3),v_5=(-1,-1,-1,-1).
\end{split}
\ee
The integral points in dual polyhedra $\nabla_4$ are
\be
\begin{split}
&v^*_0=(0,0,0,0), v^*_1=(-1,-2,-2,-2), v^*_2=(1,0,0,0), v^*_3=(0,1,0,0),\\
&v^*_4=(0,0,1,0), v^*_5=(0,0,0,1),v^*_6=(0,-1,-1,-1),
\end{split}
\ee
leading to the defining polynomial for the toric hypersurface
\be
P=a_1x_1^8+a_2x_2^8+a_3x_3^4+a_4x_4^4+a_5x_5^4+a_0x_1x_2x_3x_4x_5+a_6x_1^4x_2^4.
\ee
\subsubsection{Parallel D-branes phase}
We choose a reducible divisor $\mathcal{D}=\mathcal{D}_1+\mathcal{D}_2$ standing for the two parallel D-branes. It is realized by the degree 8 homogeneous equation
\be
\begin{split}
Q=&b_0(x_1x_2x_3x_4x_5)^2+b_1x_3^5x_1x_2x_4x_5+b_2x_3^8\\
 &\sim\prod_{i=1}^2(\phi_ia_0x_1x_2x_3x_4x_5+a_3x_3^4).
\end{split}
\ee
The vertices in the enhanced polyhedron $\tilde{\nabla}_5$ of the open-closed system are
\be
\begin{split}{\label{V3}}
&\tilde{v}_0^*=(0,0,0,0,0), \tilde{v}_1^*=(-1,-2,-2,-2,0), \tilde{v}_2^*=(1,0,0,0,0),\\
&\tilde{v}_3^*=(0,1,0,0,0), \tilde{v}_4^*=(0,0,1,0,0), \tilde{v}_5^*=(0,0,0,1,0),\\
&\tilde{v}_{6}^*=(0,-1,-1,-1,0),\tilde{v}^*_7=(0,0,0,0,1),\tilde{v}^*_8=(0,1,0,0,1),\\
&\tilde{v}_9^*=(0,2,0,0,1).
\end{split}
\ee
Together with the compactifying point $\tilde{v}_{c}^*$, these vertices describe the polyhedron $\nabla_5$ associating to F-theory compactification $W_4$.

The generators of Mori cone are
\begin{equation}{\label{L3}}
\begin{array}{ccccccccccccccccccccccccc}
     & &  & 0 &1&2&3&4&5&6&7&8&9&c & \\
  l^1&=&( & -3&0&0&0&1&1&1&-1&1&0&0 &)\\
  l^2&=&( & 0 &1&1&0&0&0&-2&0&0&0&0 &)\\
  l^3&=&( &0&0&0&0&0&0&0&1&-2&1&0 &)\\
  l^4&=&( &-1&0&0&1&0&0&0&0&1&-1&0 &)\\
  l^5&=&( &0&0&0&-2&0&0&0&0&0&1&1 &).
\end{array}
\end{equation}
The K\"{a}hler form is $J=\sum_ak_aJ_a$ for $a=\{1,2,3,4,5\}$, where $J_a$ denotes the basis of $H^{1,1}(W_4)$ dual to the Mori cone generated by $l^a$ in (\ref{L3}). And $k_a$ are flat coordinates on the K\"{a}hler moduli space of the mirror four-fold $W_4$.
Then we choose basis elements (\ref{basis321}) of $H_4(W_4)$ which are defined by intersections of the toric divisors $D_i$ corresponding to the $\tilde{v}_i^*$.
\be\label{basis321}
\begin{split}
&\gamma_1=D_3\cap D_6,~\gamma_2=D_3\cap D_1,~\gamma_3=D_6\cap D_c,~\gamma_4=D_2\cap D_c,\\
&\gamma_5=D_2\cap D_7,~\gamma_6=D_6\cap D_7,~\gamma_7=D_6\cap D_8,~\gamma_8=D_1\cap D_8
\end{split}
\ee
Selecting three interesting 4-cycles, namely $\gamma_4,~\gamma_5$ and $\gamma_5+\gamma_8$, in $H_4(W_4,{Z})$, the bulk potential and the superpotentials for the parallel branes are recovered.
After changing the variables to visualize the closed $(t)$ and open $(\hat{t})$ moduli as follows
\be
t_1=k_1+k_3+k_4,~~t_2=k_2,~~\hat{t}_1=k_3+k_4,~~\hat{t}_2=k_4,
\ee
the leading terms corresponding to them are
\be\label{leading_term3}
\tilde{\Pi}^*_{2,1}=2t_1^2,~~\tilde{\Pi}^*_{2,2}=\frac{3}{2}(t_1-\hat{t}_1)^2,~~\tilde{\Pi}^*_{2,3}=\frac{3}{2}(t_1-\hat{t}_2)^2.
\ee

Similar to the one closed moduli case, the $\tilde{\Pi}^*_{2,1}$ depends only on the close moduli $t_1$ which leading the prepotential function $F_{t}(t)$ and the $\tilde{\Pi}^*_{2,2},~\tilde{\Pi}^*_{2,3}$ leading the D-brane superpotentials are equipped with the ${Z}_2$ symmetry with respect to the open moduli $\hat{t}_1$ and $\hat{t}_2$, which is inherited from the symmetry of two parallel branes.

The instanton corrections of the period integrals are recorded in the solution of the generalized GKZ system corresponding to the enhanced polyhedron $\tilde{\nabla}_5$. We identify the bulk potential and the superpotentials with the exact solutions to the GKZ system which lead by $\tilde{\Pi}^*_{2,1}, \tilde{\Pi}^*_{2,2}$ and $\tilde{\Pi}^*_{2,3}$ respectively.
In algebraic coordinates (\ref{ac})
\be
z_1=\frac{a_4a_5a_6b_1}{a_0^3b_0},~~~z_2=\frac{a_1a_2}{a_6^2},~~~z_3=\frac{b_0b_2}{b_1^2},~~~z_4=\frac{a_3b_1}{a_0b_2},
\ee
the fundamental period and the logarithmic periods,
\begin{equation}
\Pi_0(z)=w_0(z;0),~\Pi_{1,i}(z)=\partial_{\rho_i}w_0(z;\rho)|_{\rho_i=0},~\Pi_{2,n}(z)=\sum_{i,j}K_{i,j;n}\partial_{\rho_i}\partial_{\rho_j}w_0(z;\rho)|_{\rho=0},
\end{equation}
solve the generalized GKZ system governed by charge vectors (\ref{L3}).
The flat coordinates are given by
\be
k_i=\frac{\Pi_{1,i}(z.)}{\Pi_0(z.)}=\frac{1}{2\pi i}log~z_i+...~.
\ee
Then the mixed inverse mirror maps in terms of $q_i=exp(2\pi ik_i)$ for $\{i=1,2,3,4\}$ are
\be\label{mm3}
\begin{split}
z_1=&{q_1} + 6q_1^2 + {q_1}{q_2} + 12q_1^2{q_2} + 6q_1^2q_2^2 + {q_1}{q_3} + {q_1}{q_2}{q_3} + 6q_1^2q_3^2 - 2{q_1}{q_3}{q_4} - 84q_1^2{q_3}{q_4} - ...\\
z_2=&{q_2} - 2q_2^2 - 48{q_1}{q_2}{q_3}{q_4} + 240{q_1}q_2^2{q_3}{q_4} - 264q_1^2{q_2}q_3^2q_4^2 - 5856q_1^2q_2^2q_3^2q_4^2+...\\
z_3=&{q_3} - 6{q_1}{q_3}  - 6{q_1}{q_2}{q_3} + 144q_1^2{q_2}{q_3} + 27q_1^2q_2^2{q_3} - 2q_3^2 + 30{q_1}q_3^2 - 216q_1^2q_3^2 + 30{q_1}{q_2}q_3^2  - \\
&{q_3}{q_4} + 6{q_1}{q_3}{q_4} - 27q_1^2{q_3}{q_4} + 6{q_1}{q_2}{q_3}{q_4} + 5q_3^2{q_4} - 60{q_1}q_3^2{q_4} - 3q_3^2q_4^2 +...\\
z_4=&{q_4} + {q_3}{q_4} - 12{q_1}{q_3}{q_4} + 27q_1^2{q_3}{q_4} - 12{q_1}{q_2}{q_3}{q_4} + q_4^2 - 32{q_1}{q_3}q_4^2  + q_3^2q_4^2 +...~.
\end{split}
\ee
According to the leading terms (\ref{leading_term3}), we find the relative periods which correspond to the closed-string period and D-brane superpotentials in the A-model as follows:
\be\label{spp2}
\begin{split}
F_{t}(t)\equiv\Pi_{2,1}=&2t_1^2+ \frac{1}{4\pi^2}(4{q_2} + {q_2^2} + \frac{{4q_2^3}}{9} + 640{q_1}q_2 + q_1^2( {72224q_2 + 20224q_2^2} ) + \\
&q_1^3( {753920{q_2} + 15078400q_2^2 + \frac{{7787008}}{9}q_2^3})+...)\\
\mathcal{W}_1(t,\hat{t}_1)\equiv\Pi_{2,2}=&\frac{3}{2}(t_1-\hat{t}_1)^2+ \frac{1}{4\pi^2}(3{q_2} + \frac{3{q_2^2}}{4} + 12\hat{q}_1 + 3\hat{q}_1^2  - 27{q_1}\hat{q}_1^{-1} + 66\hat{q}_1{q_1}  -\\
&63{q_2}{q_1}\hat{q}_1^{-1} + 360{q_2}{q_1} + 282\hat{q}_1{q_2}{q_1}   + \frac{{405}}{4}{q_1^2}\hat{q}_1^{-2} - 324{q_1^2}\hat{q}_1^{-1} + \\
&11862q_2^2{q_1^2} - 9288{q_2}{q_1^2}\hat{q}_1^{-1} + 39024{q_2}{q_1^2}+... )\\
\mathcal{W}_2(t,\hat{t}_2)\equiv\Pi_{2,3}=&\frac{3}{2}(t_1-\hat{t}_2)^2+ \frac{1}{4\pi^2}(3{q_2} + \frac{3{q_2^2}}{4} + 12\hat{q}_2 + 3\hat{q}_2^2  - 27{q_1}\hat{q}_2^{-1} + 66\hat{q}_2{q_1}  -\\
&63{q_2}{q_1}\hat{q}_2^{-1} + 360{q_2}{q_1} + 282\hat{q}_2{q_2}{q_1}   + \frac{{405}}{4}{q_1^2}\hat{q}_2^{-2} - 324{q_1^2}\hat{q}_2^{-1} + \\
&11862q_2^2{q_1^2}  - 9288{q_2}{q_1^2}\hat{q}_2^{-1} + 39024{q_2}{q_1^2}+... ),
\end{split}
\ee
where $q_1=exp(2\pi it_1)$, $q_2=exp(2\pi it_2)$, $\hat{q}_1=exp(2\pi i\hat{t}_1)$ and $\hat{q}_2=exp(2\pi i\hat{t}_2)$. The closed-string period $F_{t}(t)$ only depends on the closed modulus $t_1$ and $t_2$. It is noticeable that even though the classical term is contributed by one close-string deformation only, the instanton corrections are received from both of two bulk parameters. The similar phenomenon emerges in the D-brane superpotentials $\mathcal{W}_1(t,\hat{t}_1)$, and $ \mathcal{W}_2(t,\hat{t}_2)$. They are leaded by the classical part which relies on $t_1$, but $t_2$ is observed in the quantum part. This phenomenon may not be noticed in a computation using the method of subsystem integral, since the calculation of leading terms would not be involved. Moreover, the D-brane superpotentials $\mathcal{W}_1(t,\hat{t}_1), \mathcal{W}_2(t,\hat{t}_2)$ show $Z_2$ symmetry with respect to $\hat{t}_1$ and $\hat{t}_2$ as what we observed in the last model.

Also we list the superpotential for the D-brane with one open-string modulus described by the divisor $\mathcal{D}_0=\phi_0a_0x_1x_2x_3x_4x_5+a_3x_3^4$ as follows:
\be
\begin{split}\label{w02}
\mathcal{W}_0(t,\hat{t}_0)=&\frac{3}{2}(t_1-\hat{t}_0)^2+ \frac{1}{4\pi^2}(3{q_2} + \frac{3{q_2^2}}{4} + 12\hat{q}_0 + 3\hat{q}_0^2  - 27{q_1}\hat{q}_0^{-1} + 66\hat{q}_0{q_1}  -\\
&63{q_2}{q_1}\hat{q}_0^{-1} + 360{q_2}{q_1} + 282\hat{q}_0{q_2}{q_1}   + \frac{{405}}{4}{q_1^2}\hat{q}_0^{-2} - 324{q_1^2}\hat{q}_0^{-1} + \\
&11862q_2^2{q_1^2}  - 9288{q_2}{q_1^2}\hat{q}_0^{-1} + 39024{q_2}{q_1^2}+... ),
\end{split}
\ee
where $\hat{t}_0$ is the only open modulus and $\hat{q}_0=exp(2\pi i\hat{t}_0)$.
Notice that the $\mathcal{W}_1(t, \hat{t}_1)$ and $\mathcal{W}_2(t, \hat{t}_2)$ (\ref{spp2}) are in agreement with the superpotential $\mathcal{W}_0(t, \hat{t}_0)$ (\ref{w02}) when $\hat{t}_0=\hat{t}_1=\hat{t}_2$, namely
\be\label{rlt2}
\mathcal{W}_{1}(t,\hat{t}_{1})=\mathcal{W}_{2}(t,\hat{t}_{2})=\mathcal{W}_0(t,\hat{t}_0).
\ee

\subsubsection{Coincident D-branes phase}
After constructing the enhanced polyhedron $\nabla_5$, the defining polynomial $\tilde{P}$ of the dual B-model 4-fold $M_4$ is written down as follows:
\be
\begin{split}
\tilde{P}=&a_1x_1^8x_6^6+a_2x_2^8x_9^6+a_3x_3^4x_6x_7x_8x_9x_{10}^4+a_4x_4^4x_8^3+a_5x_5^4x_7^3+a_6x_1^4x_2^4x_6^3x_9^3+a_7x_1^2x_2^2x_3^2x_4^2x_5^2+\\
&a_8x_1x_2x_3^5x_4x_5x_{10}^3+a_9x_3^8x_{10}^6+a_{10}x_6^2x_7^2x_8^2x_9^2x_{10}^2+a_0x_1x_2x_3x_4x_5x_6x_7x_8x_9x_{10}.
\end{split}
\ee
We denote the coefficients of the monomials in the polynomial as $a_i$'s, and the relations between the notations in the D-brane geometry and the 4-fold are
\be
a_i=\begin{cases}
a_i&0 \leq i\leq 6,\\
b_{i-6}&
7\leq i \leq 9,\\
c&
i=10.
\end{cases}
\ee
When two D-branes coincide, the defining equation for D-branes becomes:
\be
Q\sim(\phi a_0x_1x_2x_3x_4x_5+a_3x_3^4)^2,
\ee
which means the position parameters of two individual D-branes are equal, $\phi_1=\phi_2=\phi$. Correspondingly, the equivalent condition is
\be\label{coincide2}
a_8^2=4a_7a_9
\ee
giving rise to the perfect square $(a_0x_2x_3x_4x_5\pm x_3^4)^2$ in the $\tilde{P}$. Obviously it gives rise to the singular curve on $M_4$ and the dual description of the coincident on the A-model side is the blow-down of the exceptional divisor developing the curve singularity on the $W_4$.

\begin{figure}[htbp]
  \centering
  \includegraphics[width=0.65\textwidth]{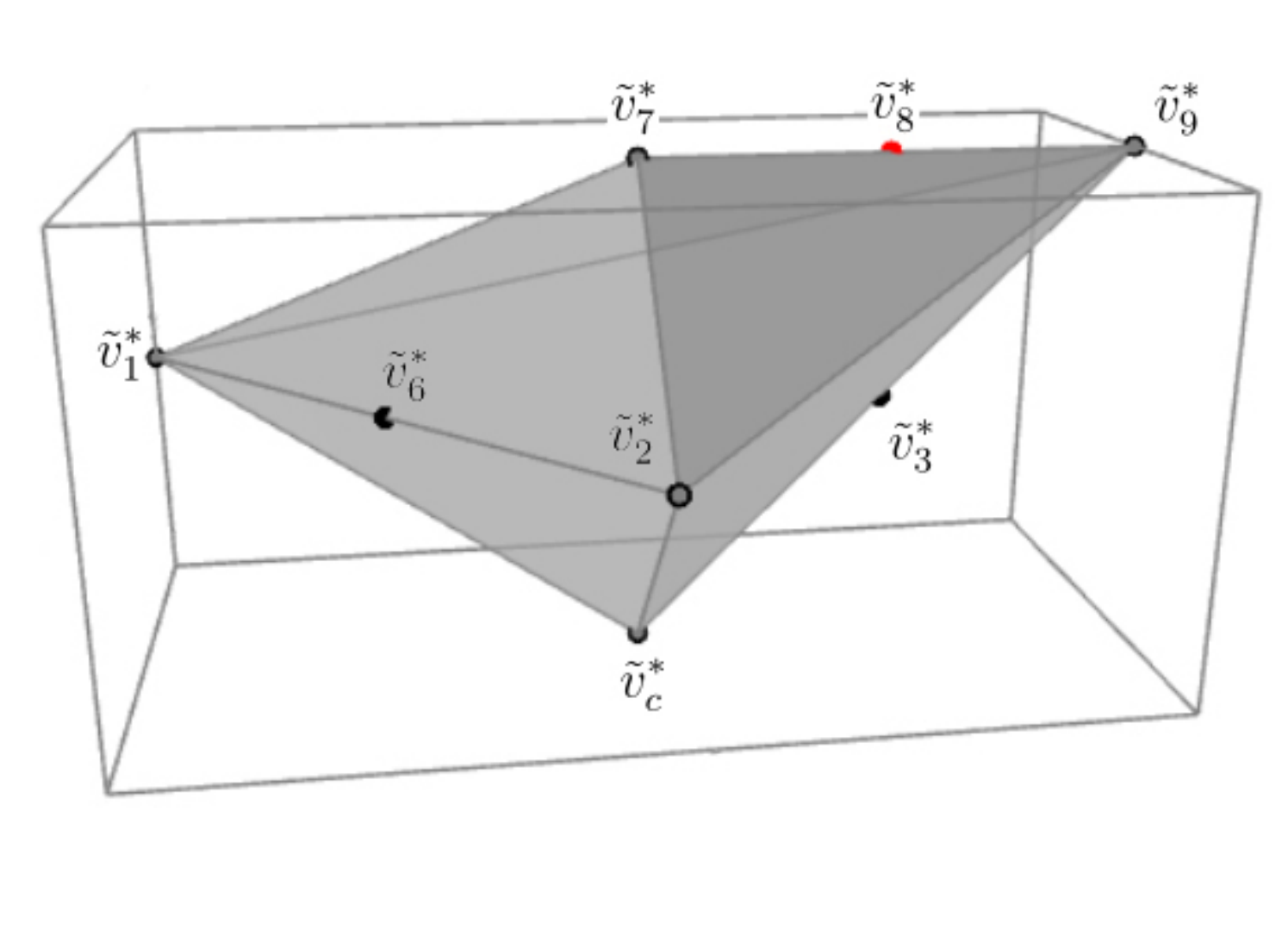}
  \caption{projected polyhedra of the $n=2$ coincident D-brane in the $X_8(1,1,2,2,2)$}\label{fig:5}
\end{figure}

The polyhedra of the $N=2$ coincident D-brane geometry is projected in the 2 dimensional space spanned by $x_1,x_2$ and $x_5$ in the Figure~\ref{fig:5}. The points $\tilde{v}^*_7, \tilde{v}^*_8, \tilde{v}^*_9$ lie on the one-dimensional edge and we ignore the interior point $\tilde{v}^*_8$ to recover the singularity corresponding to the coincidence of the branes.

The new charge vectors $l^a$'s for the coincident D-branes phase are
\begin{equation}\label{L4}
\begin{array}{ccccccccccccccccccccccccc}
     & &  & 0 &1&2&3&4&5&6&7&9&c& \\
  l^1&=&( & -6&0&0&0&2&2&2&-1&1&0&)\\
  l^2&=&( &  0&1&1&0&0&0&-2&0&0&0&)\\
  l^3&=&( & -2&0&0&2&0&0&0&1&-1&0&)\\
  l^4&=&( &  0&0&0&-2&0&0&0&0&1&1&),
\end{array}
\end{equation}
and the $J_a$ denotes the basis of $H^{1,1}(W_4)$ dual to the Mori cone generated by (\ref{L4}). Take the notation $k_a$ as the flat coordinates on the K\"{a}hler moduli space of the mirror four-fold $W_4$.
Then we choose basis elements (\ref{basis322}) of $H_4(W_4)$ which are defined by intersections of the toric divisors $D_i$ corresponding to the $\tilde{v}_i^*$.
\be\label{basis322}
\begin{split}
&\gamma_1=D_3\cap D_6,~\gamma_2=D_3\cap D_1,~\gamma_3=D_6\cap D_c,\\
&\gamma_4=D_2\cap D_c,~\gamma_5=D_2\cap D_c,~\gamma_6=D_6\cap D_7
\end{split}
\ee
After changing the variables as follows to visualize the closed and open moduli
\be
t_1=k_1+k_3,~~t_2=k_2,~~\hat{t}=k_3,
\ee
we choose two interesting elements $\gamma_4$ and $\gamma_5$ in $H_4(W_4)$  corresponding to the leading terms of the period integrals as follows:
\be\label{leading_term4}
\tilde{\Pi}^*_{2,1}=2t_1^2,~~\tilde{\Pi}^*_{2,2}=\frac{3}{2}(t_1-\hat{t})^2.
\ee
In (\ref{leading_term4}), the $\tilde{\Pi}^*_{2,1}$ depends on the closed modulus $t$ purely leading the bulk potential, and the $\tilde{\Pi}^*_{2,2}$ rely on closed ($t$) and open ($\hat{t}$) modulus leading the superpotential. It is noticeable that there is only one open modulus in the coincident branes phase, since the coincident condition reduces the degree of freedom of the open-closed parameter space. The open deformation parameter $\hat{t}$ describes the configuration of the two coincident D-branes.

The instanton corrections of the period integrals are recorded in the solution of the generalized GKZ system corresponding to the enhanced polyhedron $\tilde{\nabla}_5$. We identify the bulk potential and the superpotentials with the exact solutions to the GKZ system which lead by $\tilde{\Pi}^*_{2,1}$ and $\tilde{\Pi}^*_{2,2}$ respectively.
The algebraic coordinates are
\be
z_1=\frac{a_4^2a_5^2a_6^2a_9}{a_0^6a_7},~~~z_2=\frac{a_1a_2}{a_6^2},~~~z_3=\frac{a_3^2a_7}{a_0^2a_9}.
\ee
The fundamental period and the logarithmic periods,
\begin{equation}
\Pi_0(z)=w_0(z;0),~\Pi_{1,i}(z)=\partial_{\rho_i}w_0(z;\rho)|_{\rho_i=0},~\Pi_{2,n}(z)=\sum_{i,j}K_{i,j;n}\partial_{\rho_i}\partial_{\rho_j}w_0(z;\rho)|_{\rho=0},
\end{equation}
solve the generalized GKZ system governed by charge vectors (\ref{L4}).
The flat coordinates are
\be
k_i=\frac{\Pi_{1,i}(z)}{\Pi_0(z)}=\frac{1}{2\pi i}log~z_i+...~.
\ee
Then the mixed inverse mirror maps in terms of $q_i=exp(2\pi ik_i)$ for $\{i=1,2,3\}$ are
\be
\begin{split}
z_1=&{q_1} + 90q_1^2 + 2{q_1}{q_2} + 540q_1^2{q_2} + {q_1}q_2^2 + 900q_1^2q_2^2 - {q_1}{q_3} - 18504q_1^2{q_3} - 2{q_1}{q_2}{q_3} - \\
&146304q_1^2{q_2}{q_3} - {q_1}q_2^2{q_3} - 255600q_1^2q_2^2{q_3} + 8172q_1^2q_3^2 + 84312q_1^2{q_2}q_3^2 + 152280q_1^2q_2^2q_3^2 + ...\\
z_2=&{q_2} - 2q_2^2 - 7560{q_1}{q_2}{q_3} + 30240{q_1}q_2^2{q_3} - 29521380q_1^2{q_2}q_3^2 - 706834800q_1^2q_2^2q_3^2 +...\\
z_3=&{q_3} - 90{q_1}{q_3} + 13275q_1^2{q_3} - 360{q_1}{q_2}{q_3} + 244800q_1^2{q_2}{q_3} - 90{q_1}q_2^2{q_3} + 550800q_1^2q_2^2{q_3} + \\
&q_3^2 - 6228{q_1}q_3^2 - 716670q_1^2q_3^2 - 24912{q_1}{q_2}q_3^2 - 26422560q_1^2{q_2}q_3^2 - 6228{q_1}q_2^2q_3^2 - \\
&59677560q_1^2q_2^2q_3^2 +...~.
\end{split}
\ee
Matching the classical terms $\tilde{\Pi}^*_{2,1}$ and $\tilde{\Pi}^*_{2,2}$ of the bulk potential and the superpotentials with the exact solutions to the GKZ system, we obtain
\be
\begin{split}
F_t(t)=&2t_1^2 + \frac{1}{4\pi^2}(86528{\check{q}_1} + 16{\check{q}_2} + \frac{{369061888\check{q}_1^2}}{3}  + 1010816{\check{q}_1}{\check{q}_2}+ 4\check{q}_2^2 + \frac{{114663991360}}{3}\check{q}_1^2{\check{q}_2}  - \\
&574976{\check{q}_1}\check{q}_2^2 + 134972028960\check{q}_1^2\check{q}_2^2...)\\
\mathcal{W}_c(t,\hat{t})=&\frac{3}{2}(t_1-\hat{t})^2 +\frac{1}{4\pi^2}(- 1269{q_1} + \frac{{662805q_1^2}}{4} + 12{q_2} - 7236{q_1}{q_2} + 4299300q_1^2{q_2} + 3q_2^2 - \\
&2889{q_1}q_2^2 + 11682225q_1^2q_2^2 + 30{q_3} + 49518{q_1}{q_3} - 25407684q_1^2{q_3} + 551448{q_1}{q_2}{q_3} - \\
&680827392q_1^2{q_2}{q_3} + 306990{q_1}q_2^2{q_3} - 1835666712q_1^2q_2^2{q_3} + 9q_3^2 + 228798{q_1}q_3^2 + \\
&58092030q_1^2q_3^2 + 1308312{q_1}{q_2}q_3^2 + 20224177488q_1^2{q_2}q_3^2 + 523638{q_1}q_2^2q_3^2 + \\
&71930832408q_1^2q_2^2q_3^2 +...),
\end{split}
\ee
where $\check{q}_1 = exp(2\pi it_1)$ and $\check{q}_2=exp(2\pi it_2)$.

\subsubsection{The disk invariants of two phases}
Similar to the superpotentials of D-branes on the mirror quintic, the $Z_2$ symmetry of the superpotentials $\mathcal{W}_1(t,\hat{t}_1)$ and $\mathcal{W}_2(t,\hat{t}_1)$ appears again in the parallel D-brane phase of the D-brane system, D-branes on the hypersurface $X_8(1,1,2,2,2)$.
In this model, the $\mathcal{W}_1(t, \hat{t}_1)$ and $\mathcal{W}_2(t, \hat{t}_2)$ (\ref{spp2}) are contributed by the two parallel D-branes which are parameterized by the open modulus $\hat{t}_1$ and $\hat{t}_2$ respectively.
When $\hat{t}_0=\hat{t}_1=\hat{t}_2$ they equal to the superpotential $\mathcal{W}_0(t, \hat{t}_0)$ (\ref{w02}) of the D-brane system which contains only one D-brane as Eq.(\ref{rlt2}).
It means in the parallel phase of this model there is no interaction between the two parallel topological D-branes too.
The $U(1)$ Oogrui-Vafa invariants extracted from the superpotential $\mathcal{W}_1(t,\hat{t}_1)$ are listed in the Table~\ref{tab:3} and the superpotential $\mathcal{W}_1(t,\hat{t}_2)$ for the other brane gives the same invariants up to a transformation of the index $n_1\rightarrow n_1,~n_2\rightarrow n_2,~n_3\rightarrow n_3=n_1,~n_4\rightarrow n_4$.

In the coincident phase, the phase transition corresponds to the blow-down of the exceptional divisors in the dual F-theory 4-fold generating the $A_1$ singularity. Meanwhile, the coincidence of the two D-branes give rise to an enhancement of the gauge group $U(1)\times U(1)\rightarrow SU(2)$ in terms of gauge theory on the worldvolume of the D-brane.
Table~\ref{tab:4} presents the prediction of the $SU(2)$ Ooguri-Vafa invariants which are extracted from the coincident D-brane superpotential $\mathcal{W}_c(t,\hat{t})$.

Table~\ref{tab:3} and Table~\ref{tab:4} display the first several terms of the Ooguri-Vafa invariants of the D-brane system in the parallel phase and the coincident phase respectively.
The $U(1)$ Ooguri-Vafa invariants are the same as the invariants derived from the single D-brane system, so the invariants in the Table~\ref{tab:3} also stand for the disk invariants of the single D-brane system.
Although the two parallel D-branes locate on the same position and it looks like there is only one D-brane in the D-brane system in terms of set theory, the $SU(2)$ disk invariants of the coincident D-brane geometry, in the Table~\ref{tab:4}, are totally different from the data in the Table~\ref{tab:3}, as the disk invariants of the single D-brane system.

\begin{figure}[H]
  \setlength{\abovecaptionskip}{-1cm}
  \setlength{\belowcaptionskip}{-0.5cm}
  \centering
  \includegraphics[width=1\textwidth]{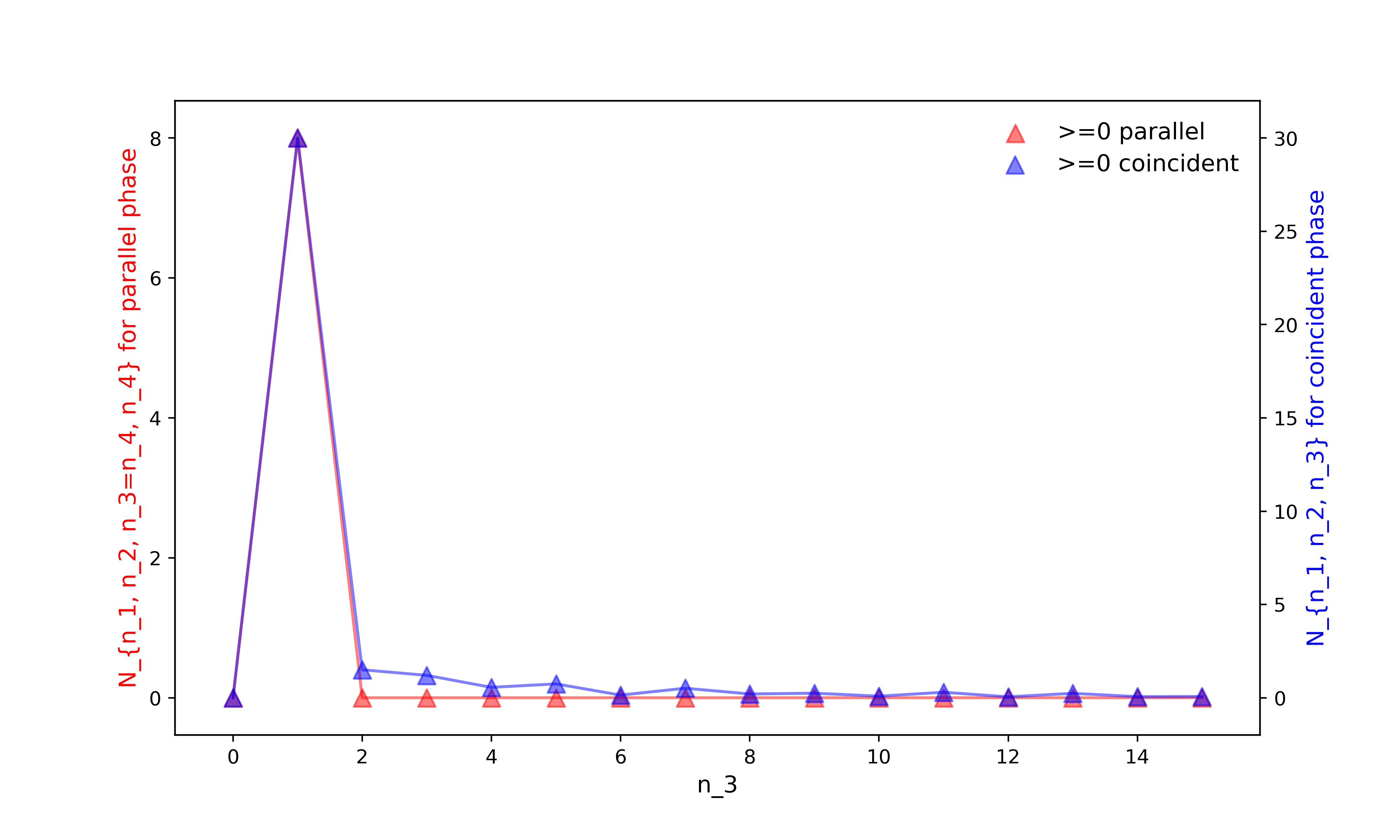}
  \caption{}\label{fig:6}
\end{figure}

\begin{figure}[H]
  \setlength{\abovecaptionskip}{-1cm}
  \setlength{\belowcaptionskip}{-0.5cm}
  \centering
  \includegraphics[width=1\textwidth]{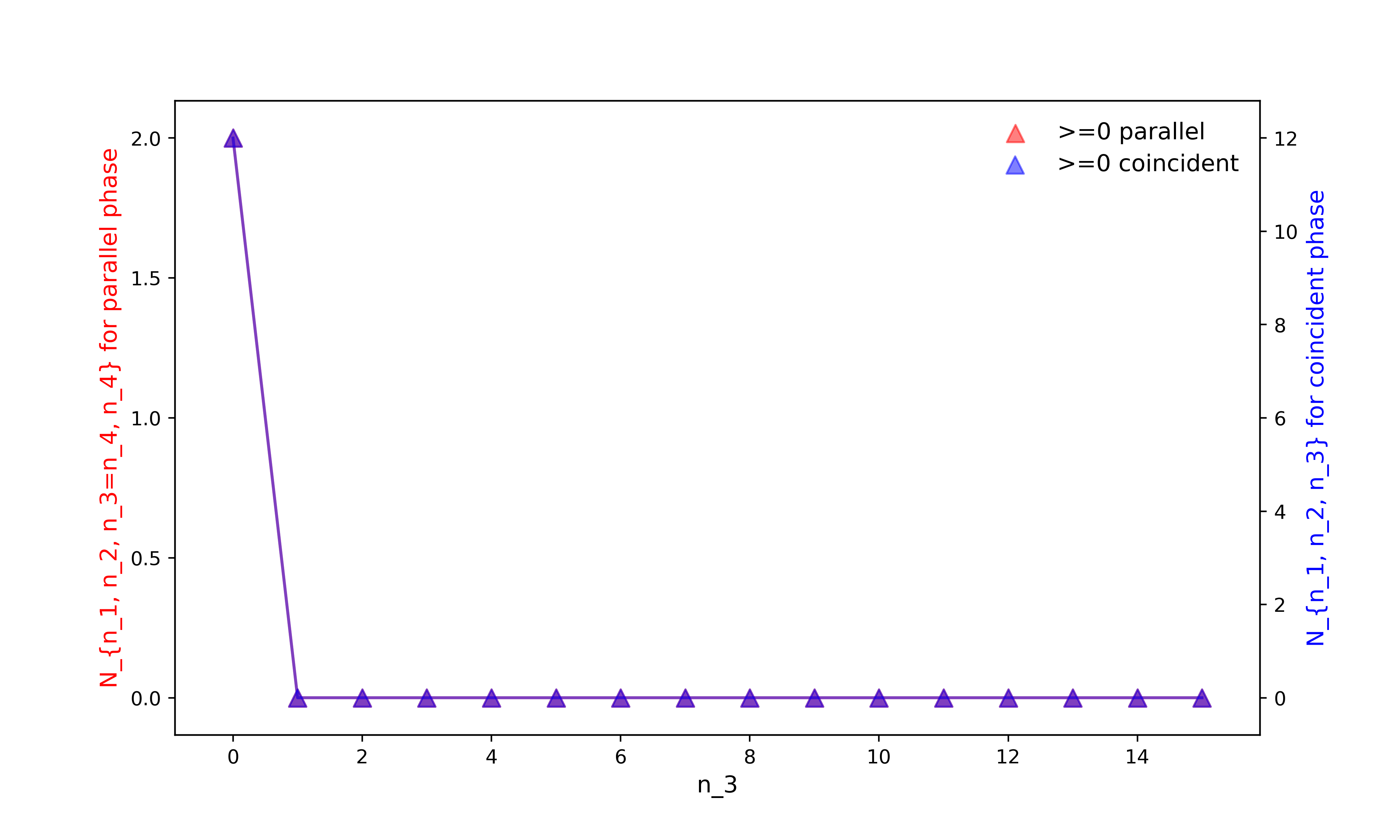}
  \caption{}\label{fig:7}
\end{figure}

\begin{figure}[H]
  \setlength{\abovecaptionskip}{-1cm}
  \setlength{\belowcaptionskip}{-0.5cm}
  \centering
  \includegraphics[width=1\textwidth]{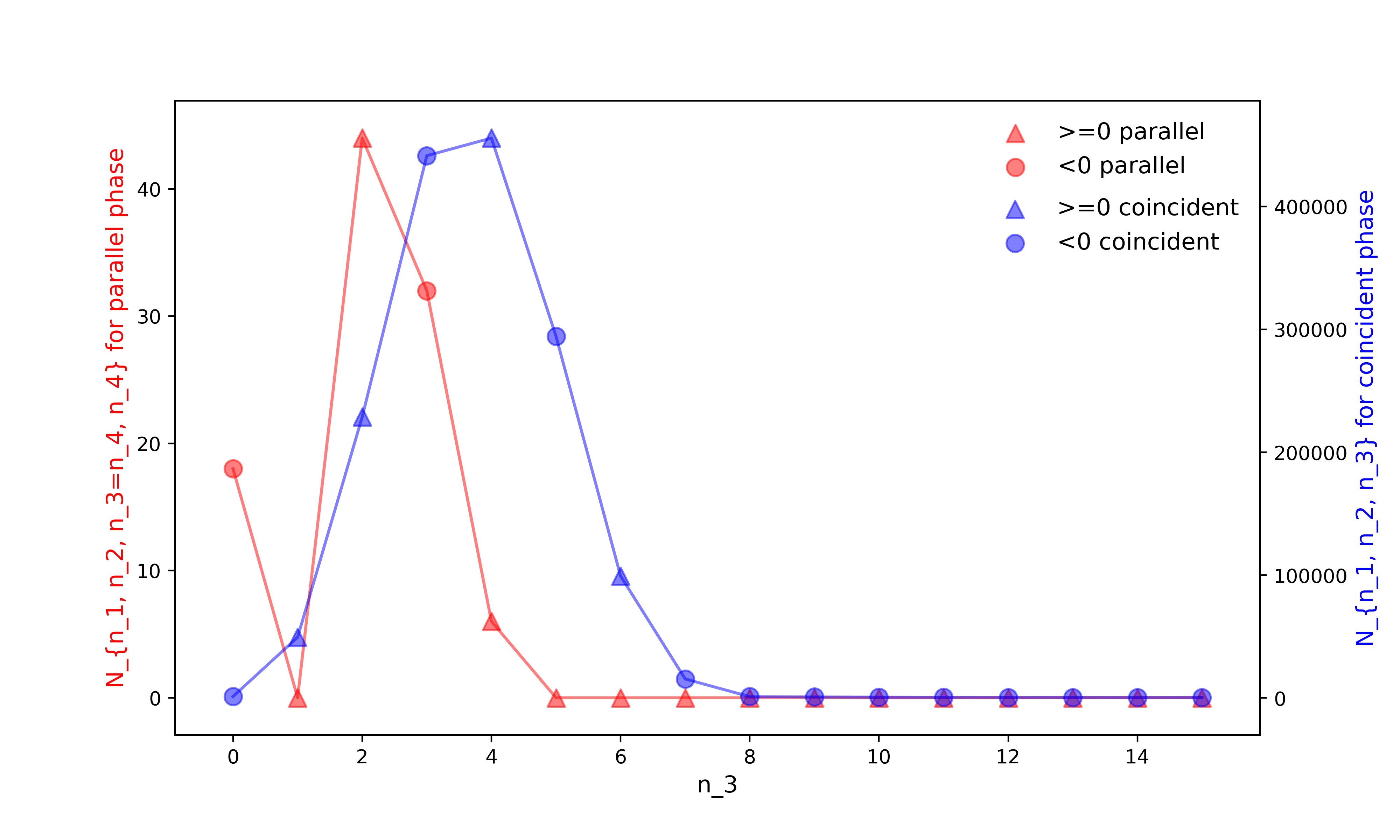}
  \caption{}\label{fig:8}
\end{figure}

\begin{figure}[H]
  \setlength{\abovecaptionskip}{-1cm}
  \setlength{\belowcaptionskip}{-0.5cm}
  \centering
  \includegraphics[width=1\textwidth]{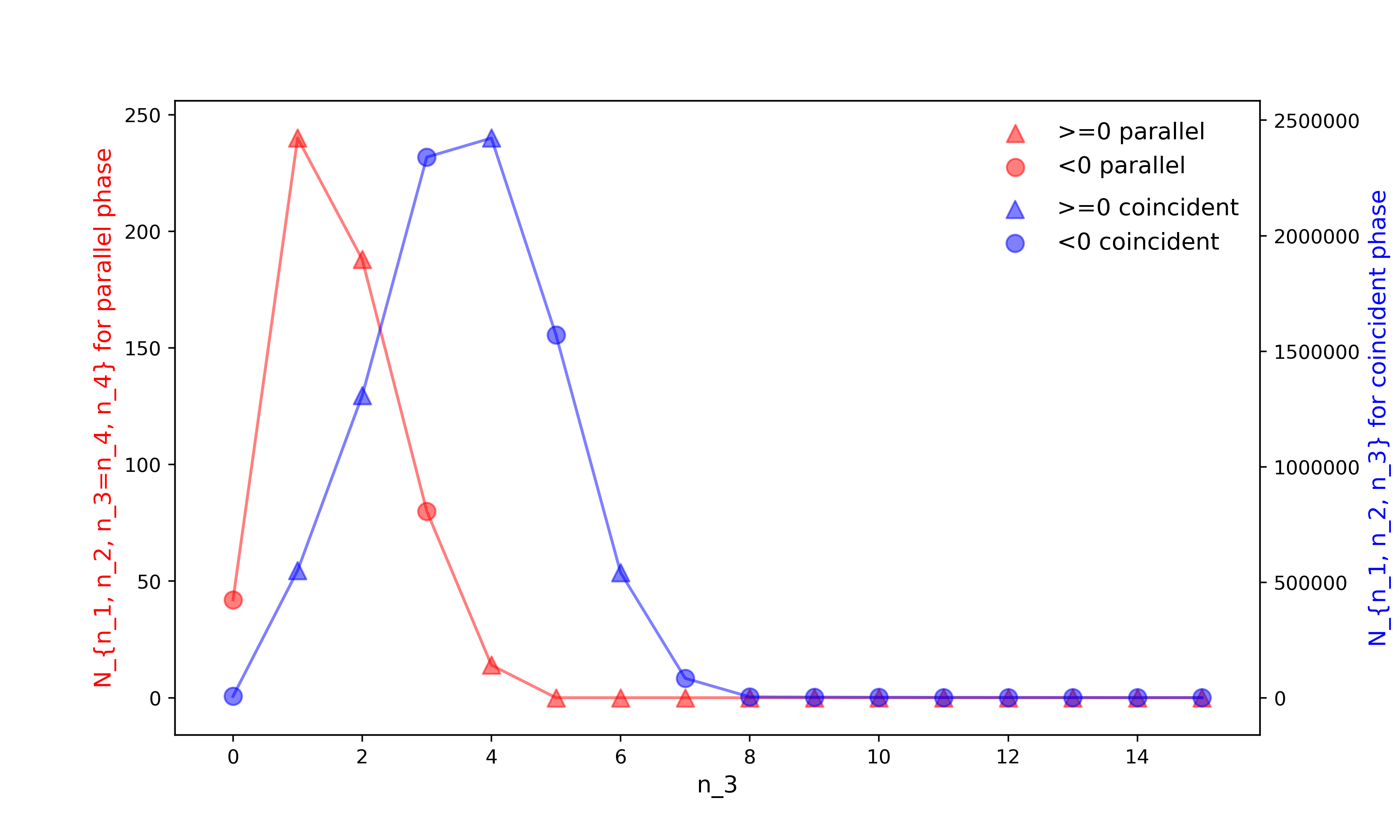}
  \caption{}\label{fig:9}
\end{figure}

In the Figure~\ref{fig:6}, Figure~\ref{fig:7}, Figure~\ref{fig:8} and Figure~\ref{fig:9}, we compare the absolute value of the Ooguri-Vafa invariants for parallel phase and the coincident phase of the D-brane system.
Similarly the Ooguri-Vafa invariants shape wave-packets when the index $n_1$ and $n_2$ has been fixed.

In the Figure~\ref{fig:6} as $n_1=0,~n_2=0$, the highest point among the red points (for parallel phase) is $N_{0,0,1,1}=8$ while the peak value for coincident phase is $N_{0,0,1}=30$. The two wave-packets are centered at $n_3=1$ both with the same duration 2.

In the Figure~\ref{fig:7} when $n_1=0,~n_2=1$, it seems only half of the wave-packets appear in the picture and the peak values are 2 and 12 for the parallel phase and the coincident phase respectively. Both of them are centered at $n_2=0$ with half of duration 1.

In the Figure~\ref{fig:8}, there is a point $N_{1,0,1,1}=0$ distorts the red wave-packet, however generally the points for invariants of two phases form the wave-packets still. The red wave-packet (parallel phase) is centered at $n_3=2$, while the blue one (coincident phase) is centered at $n_3=4$ which is 2 units away from the center of red one on the right side. And their durations are 5 (parallel phase) and 8 (coincident phase) respectively.

In the Figure~\ref{fig:9} when the $n_1$ and $n_2$ are fixed at 1, the peak value of the Ooguri-Vafa invariants for parallel phase is $N_{1,1,1,1}=240$, while the largest number in the invariants for coincident phase is $N_{1,1,4}=84793188/35$ which is over 10000 times of the $N_{1,1,1,1}$. The center of blue wave-packet shifts right 3 unit from the center of red wave-packet and the duration of coincident one (8) is wider than the parallel one (5) still.

In this model, we also find that the wave-packet for coincident phase are higher than the wave-packet for parallel phase, and the difference of hight between the wave-packets for two phases become larger as the index $n_1$ and $n_2$ increase. In all figures, with the increase of the index $n_3$, the invariants for two phases become smaller and finally equal to or near to zero.
Moreover, the wave-packet durations for the coincident phase are wider than the parallel phase and the center of the wave-packets for two phases are distinct.
These diagrams reflect the spectrum of BPS states for the two different phases of the D-brane system. As we mentioned in the last model, the coincidence of the D-branes generates much more new states than the situation of parallel which means more complicate structure of spectrum and larger number of states at energy level in the spectrum.
It suggests the emergence of the phase transition in this model.

\subsection{D-branes on hypersurface $X_9(1,1,1,3,3)$}
In this model, the ambient toric variety $P_{\Sigma(\Delta_4)}$ is determined by the lattice points in the polyhedron $\Delta_4$ as follows
\be
\begin{split}
&v_1=(3,-6,1,1), v_2=(0,0,-2,1), v_3=(-6,3,1,1),\\
&v_4=(3,3,1,1),v_5=(0,0,1,-2).
\end{split}
\ee
The integral points in dual polyhedra $\nabla_4$ are
\be
\begin{split}
&v^*_0=(0,0,0,0), v^*_1=(0,0,0,-1), v^*_2=(0,0,-1,0), v^*_3=(0,0,1,1),\\
&v^*_4=(-1,0,1,1), v^*_5=(0,-1,1,1),v^*_6=(1,1,1,1).
\end{split}
\ee
Then the defining polynomial for the toric hypersurface $M_3$ is
\be
P=a_1x_1^3+a_2x_2^3+a_3x_3^3x_4^3x_5^3+a_4x_3^9+a_5x_4^9+a_6x_5^9+a_0x_1x_2x_3x_4x_5.
\ee
\subsubsection{Parallel D-branes phase}
We realize the two parallel D-branes by the degree 8 homogeneous equation,
\be
\begin{split}
Q=&b_0(x_1x_2x_3x_4x_5)^2+b_1x_1^4x_2x_3x_4x_5+b_2x_1^6\\
 &\sim\prod_{i=1}^2(\phi_ia_0x_1x_2x_3x_4x_5+a_1x_1^3),
\end{split}
\ee
which defines the reducible divisor $\mathcal{D}=\mathcal{D}_1+\mathcal{D}_2$ with the intersection of zero locus of $P$ and $Q$.
The enhanced polyhedron $\tilde{\nabla}_5$ corresponding to the open-closed system is described by vertices
\be
\begin{split}{\label{V5}}
&\tilde{v}_0^*=(0,0,0,0,0), \tilde{v}_1^*=(0,0,0,-1,0), \tilde{v}_2^*=(0,0,-1,0,0),\\
&\tilde{v}_3^*=(0,0,1,1,0), \tilde{v}_4^*=(-1,0,1,1,0), \tilde{v}_5^*=(0,-1,1,1,0),\\
&\tilde{v}_{6}^*=(1,1,1,1,0),\tilde{v}^*_7=(0,0,0,0,1),\tilde{v}^*_8=(0,0,0,-1,1),\\
&\tilde{v}_9^*=(0,0,0,-2,1).
\end{split}
\ee
Then the vertices of polyhedron $\nabla_5$ associating to the honest F-theory compactification $W_4$ are given by (\ref{V5}) and $\tilde{v}_c^*$.

Correspondingly, the generators of Mori cone are
\begin{equation}{\label{L5}}
\begin{array}{ccccccccccccccccccccccccc}
     & &  & 0 &1&2&3&4&5&6&7&8&9&c & \\
  l^1&=&( & -2&0&1&1&0&0&0&-1&1&0&0 &)\\
  l^2&=&( & 0 &0&0&-3&1&1&1&0&0&0&0 &)\\
  l^3&=&( &0&0&0&0&0&0&0&1&-2&1&0 &)\\
  l^4&=&( &-1&1&0&0&0&0&0&0&1&-1&0 &)\\
  l^5&=&( &0&-2&0&0&0&0&0&0&0&1&1 &).
\end{array}
\end{equation}
The K\"{a}hler form is written as $J=\sum_ak_aJ_a$ for $a=\{1,2,3,4,5\}$, where $J_a$ denotes the basis of $H^{1,1}(W_4)$ dual to the Mori cone generated by $l^a$ in (\ref{L5}). Thus $k_a$'s are flat coordinates on the K\"{a}hler moduli space of the mirror four-fold $W_4$.
Then we choose a set of basis (\ref{basis331}) of $H_4(W_4)$ which are defined by intersections of the toric divisors $D_i$ corresponding to the $\tilde{v}_i^*$.
\be\label{basis331}
\begin{split}
&\gamma_1=D_c\cap D_4,~\gamma_2=D_9\cap D_3,~\gamma_3=D_8\cap D_4,\\
&\gamma_4=D_7\cap D_4,~\gamma_5=D_4\cap D_1,~\gamma_6=D_4\cap D_3
\end{split}
\ee
Focus on three interesting 4-cycles, namely $\gamma_1,~2\gamma_2+3\gamma_4$ and $2\gamma_2+3\gamma_3+3\gamma_4$, in $H_4(W_4,{Z})$ to compute the leading terms of the bulk potential and the superpotentials for the parallel branes.
After changing the variables to visualize the closed $(t)$ and open $(\hat{t})$ moduli as follows
\be
t_1=k_1+k_3+k_4,~~t_2=k_2,~~\hat{t}_1=k_3+k_4,~~\hat{t}_2=k_4~,
\ee
one obtains
\be\label{leading_term5}
\tilde{\Pi}^*_{2,1}=\frac{9}{2}t_1^2+3t_1t_2,~~\tilde{\Pi}^*_{2,2}=(3t_1+t_2-3\hat{t}_1)^2,~~\tilde{\Pi}^*_{2,3}=(3t_1+t_2-3\hat{t}_2)^2,
\ee
corresponding to the three selected 4-cycles respectively.
Similar to the previous models, there is a leading term of the bulk potential function, $\tilde{\Pi}^*_{2,1}$, depends only on the close moduli $t$.
Whereas the $\tilde{\Pi}^*_{2,2},~\tilde{\Pi}^*_{2,3}$ leads the D-brane superpotentials which rely on the open-closed deformations and are equipped with the ${Z}_2$ symmetry which inherit from the two parallel branes.

Now we compute the instanton corrections of the period integrals by matching the classical terms of the potential functions, $\tilde{\Pi}^*_{2,1}, \tilde{\Pi}^*_{2,2}$ and $\tilde{\Pi}^*_{2,3}$, and the leading terms of the solutions of the generalized GKZ system corresponding to the enhanced polyhedron $\tilde{\nabla}_5$.
Following the Eq.(\ref{fw}), in terms of algebraic coordinates (\ref{ac})
\be
z_1=\frac{a_2a_3b_1}{a_0^2b_0},~~~z_2=\frac{a_4a_5a_6}{a_3^3},~~~z_3=\frac{b_0b_2}{b_1^2},~~~z_4=\frac{a_1b_1}{a_0b_2},
\ee
the fundamental period and the logarithmic periods read
\begin{equation}
\Pi_0(z)=w_0(z;0),~\Pi_{1,i}(z)=\partial_{\rho_i}w_0(z;\rho)|_{\rho_i=0},~\Pi_{2,n}(z)=\sum_{i,j}K_{i,j;n}\partial_{\rho_i}\partial_{\rho_j}w_0(z;\rho)|_{\rho=0}.
\end{equation}
The flat coordinates are
\be
k_i=\frac{\Pi_{1,i}(z.)}{\Pi_0(z.)}=\frac{1}{2\pi i}log~z_i+...~.
\ee
Then the mixed inverse mirror maps in terms of $q_i=exp(2\pi ik_i)$ for $\{i=1,2,3,4\}$ are
\be\label{mm5}
\begin{split}
z_1=&{q_1} + 2q_1^2 - 2{q_1}{q_2} - 8q_1^2{q_2} + 5{q_1}q_2^2 + 28q_1^2q_2^2 + {q_1}{q_3} - 2{q_1}{q_2}{q_3} + 5{q_1}q_2^2{q_3} + 2q_1^2q_3^2 - \\
&2{q_1}{q_3}{q_4} - 12q_1^2{q_3}{q_4} + 4{q_1}{q_2}{q_3}{q_4} + ...\\
z_2=&{q_2} + 6q_2^2 - 18{q_1}{q_2}{q_3}{q_4} - 234{q_1}q_2^2{q_3}{q_4} + 135q_1^2{q_2}q_3^2q_4^2 + 5130q_1^2q_2^2q_3^2q_4^2+...\\
z_3=&{q_3} - 2{q_1}{q_3} + q_1^2{q_3} + 4{q_1}{q_2}{q_3} - 4q_1^2{q_2}{q_3} - 10{q_1}q_2^2{q_3} - 2q_3^2 + 10{q_1}q_3^2 - 16q_1^2q_3^2 \\
&- 20{q_1}{q_2}q_3^2 + {q_3}{q_4} + 2{q_1}{q_3}{q_4} - q_1^2{q_3}{q_4} - 4{q_1}{q_2}{q_3}{q_4} + 5q_3^2{q_4} - 20{q_1}q_3^2{q_4} - 3q_3^2q_4^2...\\
z_4=&{q_4} + {q_3}{q_4} - 4{q_1}{q_3}{q_4} + q_1^2{q_3}{q_4} + 8{q_1}{q_2}{q_3}{q_4} - 7{q_1}{q_3}q_4^2 + q_3^2q_4^2+...~.
\end{split}
\ee
According to the leading terms (\ref{leading_term5}), we find the relative periods which corresponds to the closed-string period and D-brane superpotentials in the A-model as follows:
\be\label{spp3}
\begin{split}
F_{t}(t)\equiv\Pi_{2,1}=&\frac{9}{2}t^2_1+3t_1t_2 + \frac{1}{4\pi^2}(9{q_2} - 324{q_1}{q_2} + 3645q_1^2{q_2} + 131301q_1^3{q_2} - \\
&\frac{{135q_2^2}}{4} + 1620{q_1}q_2^2 - 29889q_1^2q_2^2 + 248184q_1^3q_2^2 + 244q_2^3 \\
&- 15552{q_1}q_2^3 + 413343q_1^2q_2^3 - 5777928q_1^3q_2^3 +...)\\
\mathcal{W}_1(t,\hat{t}_1)\equiv\Pi_{2,2}=&(3t_1+t_2-3\hat{t}_1)^2+ \frac{1}{4\pi^2}(108{q_1} + 243q_1^2 - 216{q_1}{q_2} + 4860q_1^2{q_2} + \\
&540{q_1}q_2^2 - 19926q_1^2q_2^2 + {{9q_1^2}}{{\hat{q}_1^{-2}}} - {{36q_1^2{q_2}}}{{\hat{q}_1^{-2}}} + {{126q_1^2q_2^2}}{{\hat{q}_1^{-2}}}- \\
& {{36{q_1}}}{{{\hat{q}_1^{-1}}}} - {{54q_1^2}}{{{\hat{q}_1^{-1}}}} + {{72{q_1}{q_2}}}{{{\hat{q}_1^{-1}}}} - {{1728q_1^2{q_2}}}{{{\hat{q}_1^{-1}}}} - {{180{q_1}q_2^2}}{{{\hat{q}_1^{-1}}}} + \\
&{{7020q_1^2q_2^2}}{{{\hat{q}_1^{-1}}}} + 54{\hat{q}_1} + 54{q_1}{\hat{q}_1} - 108{q_1}{q_2}{\hat{q}_1} + 270{q_1}q_2^2{\hat{q}_1} + \frac{27}{2}{{\hat{q}_1^2}} +... )\\
\mathcal{W}_2(t,\hat{t}_2)\equiv\Pi_{2,3}=&(3t_1+t_2-3\hat{t}_2)^2+ \frac{1}{4\pi^2}(108{q_1} + 243q_1^2 - 216{q_1}{q_2} + 4860q_1^2{q_2} + \\
&540{q_1}q_2^2 - 19926q_1^2q_2^2 + {{9q_1^2}}{{\hat{q}_2^{-2}}} - {{36q_1^2{q_2}}}{{\hat{q}_2^{-2}}} + {{126q_1^2q_2^2}}{{\hat{q}_2^{-2}}} -\\
&{{36{q_1}}}{{{\hat{q}_2^{-1}}}} - {{54q_1^2}}{{{\hat{q}_2^{-1}}}} + {{72{q_1}{q_2}}}{{{\hat{q}_2^{-1}}}} - {{1728q_1^2{q_2}}}{{{\hat{q}_2^{-1}}}} - {{180{q_1}q_2^2}}{{{\hat{q}_2^{-1}}}} + \\
&{{7020q_1^2q_2^2}}{{{\hat{q}_2^{-1}}}} + 54{\hat{q}_2} + 54{q_1}{\hat{q}_2} - 108{q_1}{q_2}{\hat{q}_2} + 270{q_1}q_2^2{\hat{q}_2} + \frac{27}{2}{{\hat{q}_2^2}} +... ),
\end{split}
\ee
where $q_1=exp(2\pi it_1)$, $q_2=exp(2\pi it_2)$, $\hat{q}_1=exp(2\pi i\hat{t}_1)$ and $\hat{q}_2=exp(2\pi i\hat{t}_2)$.
The closed-string period $F_{t}(t)$ only depends on the closed moduli $t_1$ and $t_2$. Similar to the last two closed moduli model, even though the classical term depends on one closed string deformation only, the instanton corrections are determined by both of two bulk parameters. Analogously, D-brane superpotentials $\mathcal{W}_1(t,\hat{t}_1)$, and $ \mathcal{W}_2(t,\hat{t}_2)$ lead by the classical part which relies on $t_1(t_2)$, but $t_2(t_1)$ is observed in the quantum part. It's remarkable that the $Z_2$ symmetry of the D-brane superpotentials $\mathcal{W}_1(t,\hat{t}_1), \mathcal{W}_2(t,\hat{t}_2)$ is also found in this model.

Again we list the superpotential for the D-brane with one open deformation modulus described by the divisor $\mathcal{D}_0=\phi_0a_0x_1x_2x_3x_4x_5+a_1x_1^3$ as follows:
\be
\begin{split}\label{w03}
\mathcal{W}_0(t,\hat{t}_0)=&(3t_1+t_2-3\hat{t}_0)^2+ \frac{1}{4\pi^2}(108{q_1} + 243q_1^2 - 216{q_1}{q_2} + 4860q_1^2{q_2} + \\
&540{q_1}q_2^2 - 19926q_1^2q_2^2 + {{9q_1^2}}{{\hat{q}_0^{-2}}} - {{36q_1^2{q_2}}}{{\hat{q}_0^{-2}}} + {{126q_1^2q_2^2}}{{\hat{q}_0^{-2}}} -\\
&{{36{q_1}}}{{{\hat{q}_0^{-1}}}} - {{54q_1^2}}{{{\hat{q}_0^{-1}}}} + {{72{q_1}{q_2}}}{{{\hat{q}_0^{-1}}}} - {{1728q_1^2{q_2}}}{{{\hat{q}_0^{-1}}}} - {{180{q_1}q_2^2}}{{{\hat{q}_0^{-1}}}} + \\
&{{7020q_1^2q_2^2}}{{{\hat{q}_0^{-1}}}} + 54{\hat{q}_0} + 54{q_1}{\hat{q}_0} - 108{q_1}{q_2}{\hat{q}_0} + 270{q_1}q_2^2{\hat{q}_0} + \frac{27}{2}{{\hat{q}_0^2}} +... ),
\end{split}
\ee
where $\hat{t}_0$ is the only open modulus and $\hat{q}_0=exp(2\pi i\hat{t}_0)$.
The $\mathcal{W}_1(t, \hat{t}_1)$ and $\mathcal{W}_2(t, \hat{t}_2)$ (\ref{spp3}) are related with the superpotential $\mathcal{W}_0(t, \hat{t}_0)$ (\ref{w03}) by
\be\label{rlt3}
\mathcal{W}_{1}(t,\hat{t}_{1})=\mathcal{W}_{2}(t,\hat{t}_{2})=\mathcal{W}_0(t,\hat{t}_0),
\ee
when $\hat{t}_0=\hat{t}_1=\hat{t}_2$.

\subsubsection{Coincident D-branes phase}
Given the enhanced polyhedron $\nabla_5$, the defining polynomial of the 4-fold $M_4$ reads
\be
\begin{split}
\tilde{P}=&a_1x_1^3x_6x_7x_8x_9x_{10}^3+a_2x_2^3x_9^2+a_3x_3^3x_4^3x_5^3x_6^2x_7^2x_8^2+a_4x_3^9x_6^6+a_5x_4^9x_8^6+a_6x_5^9x_7^6+\\
&a_7x_1^2x_2^2x_3^2x_4^2x_5^2+a_8x_1^4x_2x_3x_4x_5x_{10}^2+a_9x_1^6x_{10}^4+a_{10}x_6^2x_7^2x_8^2x_9^2x_{10}^2+\\
&a_0x_1x_2x_3x_4x_5x_6x_7x_8x_9x_{10}.
\end{split}
\ee
We still denote the coefficients of the monomials in the polynomial with $a_i$'s, and the relation between the coefficients in $P$ for the D-brane geometry and them in $\tilde{P}$ for the 4-fold is as follows:
\be
a_i=\begin{cases}
a_i&0 \leq i\leq 6,\\
b_{i-6}&
7\leq i \leq 9,\\
c&
i=10.
\end{cases}
\ee
The coincidence of the D-branes is described by the equality of position parameters of different D-branes $\phi_1=\phi_2=\phi$, then the defining equation for the parallel D-branes becomes
\be
Q\sim(\phi a_0x_1x_2x_3x_4x_5+a_3x_3^3)^2.
\ee
Correspondingly, the equivalent condition
\be\label{coincide3}
a_8^2=4a_7a_9
\ee
gives rise to the perfect square $(a_0x_1x_2x_3x_4x_5\pm x_3^3)^2$ in the $\tilde{P}$.
The dual description of the coincident on the A-model side is the blow-down of the exceptional divisor developing the curve singularity on the $W_4$.

\begin{figure}[htbp]
  \centering
  \includegraphics[width=0.65\textwidth]{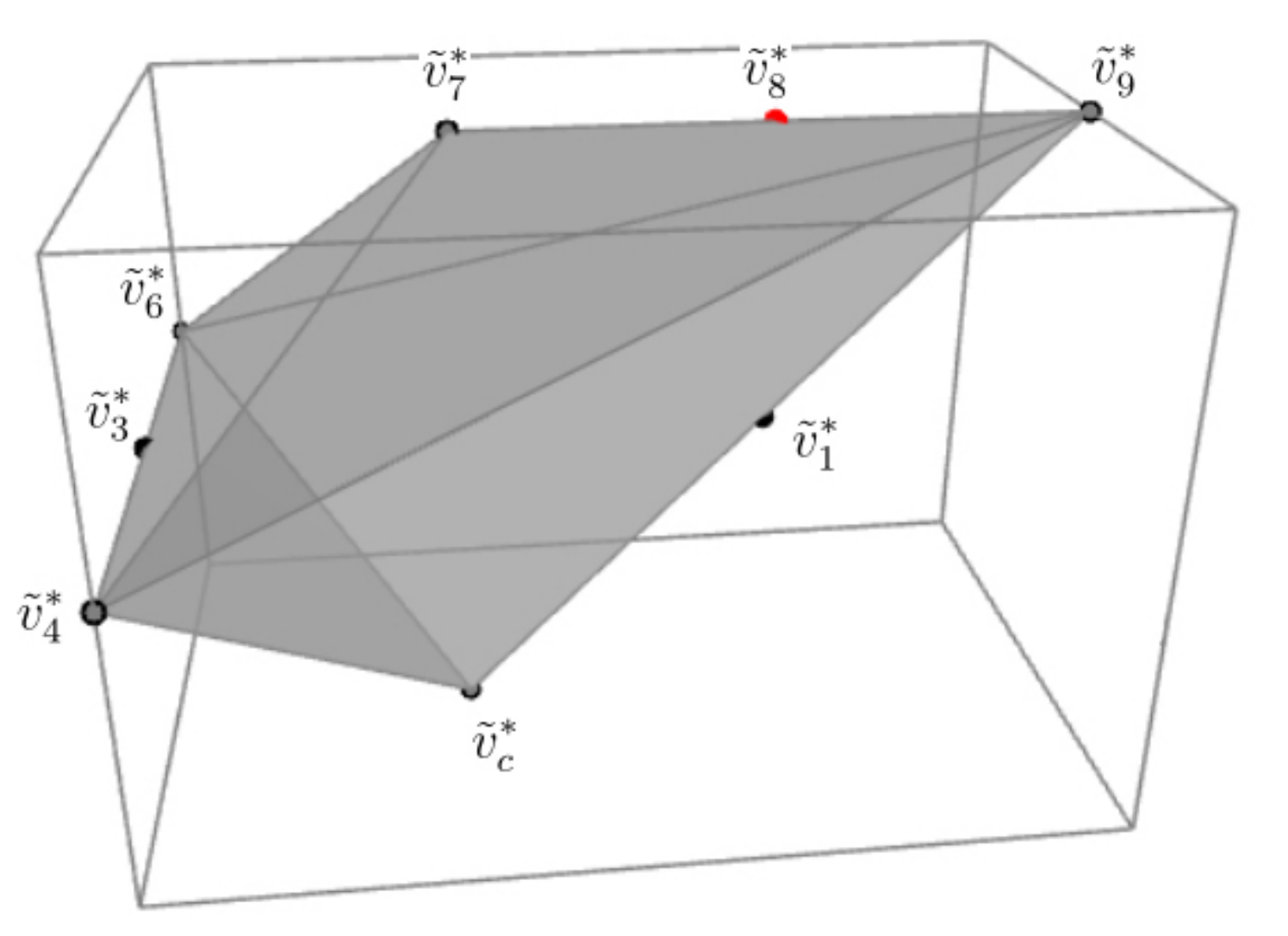}
  \caption{projected polyhedra of the $n=2$ coincident D-brane in the $X_8(1,1,1,3,3)$}\label{fig:10}
\end{figure}

The polyhedra of the $N=2$ coincident D-brane geometry is projected on the hyperplane spanned by $x_1,x_4$ and $x_5$ in the Figure~\ref{fig:10}. The points $\tilde{v}^*_7, \tilde{v}^*_8, \tilde{v}^*_9$ lie on the one-dimensional edge and we ignore the interior point $\tilde{v}^*_8$ to recover the singularity corresponding to the coincidence of the branes. Then the points configuration is
\be
\begin{split}{\label{V6}}
&\tilde{v}_0^*(0,0,0,0,0), \tilde{v}_1^*(0,0,0,-1,0), \tilde{v}_2^*(0,0,-1,0,0),\\
&\tilde{v}_3^*(0,0,1,1,0), \tilde{v}_4^*(-1,0,1,1,0), \tilde{v}_5^*(0,-1,1,1,0),\\
&\tilde{v}_{6}^*(1,1,1,1,0),\tilde{v}^*_7(0,0,0,0,1),\tilde{v}_9^*(0,0,0,-2,1).
\end{split}
\ee

The corresponding charge vectors of the coincident D-branes phase are
\begin{equation}\label{L6}
\begin{array}{ccccccccccccccccccccccccc}
     & &  & 0 &1&2&3&4&5&6&7&9&c& \\
  l^1&=&( & -4&0&2&2&0&0&0&-1&1&0&)\\
  l^2&=&( &  0&0&0&-3&1&1&1&0&0&0&)\\
  l^3&=&( & -2&2&0&0&0&0&0&1&-1&0&)\\
  l^4&=&( &  0&-2&0&0&0&0&0&0&1&1&).
\end{array}
\end{equation}
The $J_a$ are denoted as the basis of $H^{1,1}(W_4)$ dual to the Mori cone generated by (\ref{L6}). Then the corresponding local coordinates $k_a$ are flat coordinates on the K\"{a}hler moduli space of the mirror four-fold $W_4$.
We choose basis elements (\ref{basis332}) of $H_4(W_4)$ which are defined by intersections of the toric divisors $D_i$ corresponding to the $\tilde{v}_i^*$.
\be\label{basis332}
\begin{split}
&\gamma_1=D_4\cap D_c,~\gamma_2=D_1\cap D_7,~\gamma_3=D_4\cap D_7,\\
&\gamma_4=D_1\cap D_4,~\gamma_5=D_3\cap D_9
\end{split}
\ee
We study two interesting elements $\gamma_2+3\gamma_4$ and $\gamma_5$ in $H_4(W_4)$ which give rise to the leading terms of the period integrals as follows:
\be\label{leading_term6}
\tilde{\Pi}^*_{2,1}=\frac{3}{2}(3t_1+t_2)^2,~~\tilde{\Pi}^*_{2,2}=(3t_1+t_2-3\hat{t})^2,
\ee
where $t_1=k_1+k_3, t_2=k_2$, and $\hat{t}=k_3$.
It is noticeable that the $\tilde{\Pi}^*_{2,1}$ depends on the closed modulus $t_1$ and $t_2$ leading the bulk potential, and the $\tilde{\Pi}^*_{2,2}$ rely on closed ($t$) and open ($\hat{t}$) modulus leading the superpotential. The number of open modulus have been reduced to one in the coincident phase and the position of the two coincident D-branes can be described by the only open modulus.

The instanton corrections of the period integrals can be derived from the solutions to the GKZ system. We identify the bulk potential and the superpotentials with the exact solutions to the GKZ system which lead by $\tilde{\Pi}^*_{2,1}$ and $\tilde{\Pi}^*_{2,2}$ respectively.
According to (\ref{ac}), the algebraic coordinates are
\be
z_1=\frac{a_2^2a_3^2a_9}{a_0^4a_7},~~~z_2=\frac{a_4a_5a_6}{a_3^3},~~~z_3=\frac{a_1^2a_7}{a_0^2a_9}.
\ee
And the mirror map relates them with the flat coordinates on the A-model side as follows:
\be
k_i=\frac{\Pi_{1,i}(z)}{\Pi_0(z)}=\frac{1}{2\pi i}log~z_i+...~.
\ee
Then the mixed inverse mirror maps in terms of $q_i=exp(2\pi ik_i)$ for $\{i=1,2,3\}$ are
\be
\begin{split}
z_1=&{q_1} + 6q_1^2 - 4{q_1}{q_2} - 48q_1^2{q_2} + 14{q_1}q_2^2 + 264q_1^2q_2^2 - {q_1}{q_3} - 348q_1^2{q_3} + 4{q_1}{q_2}{q_3} + \\
&4404q_1^2{q_2}{q_3} - 14{q_1}q_2^2{q_3} + 183q_1^2q_3^2 +...\\
z_2=&{q_2} + 6q_2^2 - 405{q_1}{q_2}{q_3} - 3240{q_1}q_2^2{q_3} + 109665q_1^2{q_2}q_3^2 + 558810q_1^2q_2^2q_3^2 +...\\
z_3=&{q_3} - 6{q_1}{q_3} + 17q_1^2{q_3} + 24{q_1}{q_2}{q_3} + 704q_1^2{q_2}{q_3} - 84{q_1}q_2^2{q_3} - 932q_1^2q_2^2{q_3} + q_3^2 - \\
&177{q_1}q_3^2 + 468q_1^2q_3^2 + 708{q_1}{q_2}q_3^2 +...~.
\end{split}
\ee
The instanton correction of the bulk potential and the superpotentials are recovered by matching $\tilde{\Pi}^*_{2,1}$ and $\tilde{\Pi}^*_{2,2}$ with the exact solutions to the GKZ system derived by (\ref{L6}). The complete potential functions are
\be
\begin{split}
F_t(t)=&\frac{3}{2}(3t_1+t_2)^2+\frac{1}{4\pi^2}(11421{\check{q}_1} - \frac{{5772519\check{q}_1^2}}{4} + \frac{{7107420141\check{q}_1^3}}{{25}} - 80676{\check{q}_1}{\check{q}_2} + \\
&331081938\check{q}_1^2{\check{q}_2} + \frac{{1616630007798}}{{25}}\check{q}_1^3{\check{q}_2} - 54\check{q}_2^2 + 396090{\check{q}_1}\check{q}_2^2 -1534525749\check{q}_1^2\check{q}_2^2 + \\
&162\check{q}_2^3 - 2528172{\check{q}_1}\check{q}_2^3 + ...)\\
\mathcal{W}_c(t,\hat{t})=&(3t_1+t_2-3\hat{t})^2+\frac{1}{4\pi^2}( - 324q_1^2 + 216{q_1}{q_2} + 2628q_1^2{q_2} - 36q_2^2 - 1188{q_1}q_2^2 - \\
&44838q_1^2q_2^2 + 8595{q_1}{q_3} + 51534q_1^2{q_3} - 36{q_2}{q_3} - 51552{q_1}{q_2}{q_3} - 494172q_1^2{q_2}{q_3} + \\
&54q_2^2{q_3} + 243936{q_1}q_2^2{q_3} + 6296346q_1^2q_2^2{q_3} - 9q_3^2 - 8532{q_1}q_3^2 - \frac{{3529893}}{4}q_1^2q_3^2 - \\
&18{q_2}q_3^2 + 42984{q_1}{q_2}q_3^2 + 189167748q_1^2{q_2}q_3^2 + 27q_2^2q_3^2 - 211896{q_1}q_2^2q_3^2 - \\
&868393662q_1^2q_2^2q_3^2 +...),
\end{split}
\ee
where $\check{q}_1 = exp(2\pi it_1)$ and $\check{q}_2=exp(2\pi it_2)$.

\subsubsection{The disk invariants of two phases}
In the parallel phase, as we shown in the subsection 3.3.1, the superpotential $\mathcal{W}_1(t,\hat{t}_1)$ and $\mathcal{W}_2(t,\hat{t}_2)$ have the $Z_2$ symmetry with respect to exchanging open-string deformations $t_1$ and $t_2$.
Furthermore, according to the Eq.(\ref{rlt3}) when $\hat{t}_0=\hat{t}_1=\hat{t}_2$ the superpotential of the individual D-brane in the D-brane system with two parallel D-branes is identified with the superpotential $\mathcal{W}_0(t,\hat{t}_0)$ (\ref{w03}) of the only D-brane in the single D-brane system with open deformation modulus $\hat{t}_0$.
It means the superpotentials are generated by the two separated D-branes independently and it is also a signal of decoupling of the parallel D-branes in this model.
Thus $\mathcal{W}_1(t,\hat{t}_1)$ and $\mathcal{W}_2(t,\hat{t}_2)$ give the same $U(1)$ Ooguri-Vafa invariants up to an index transformation $n_1\rightarrow n_1,~n_2\rightarrow n_2,~n_3\rightarrow n_3=n_1,~n_4\rightarrow n_4$. We list the predictions of the $U(1)$ Oogrui-Vafa invariants for $\mathcal{W}_1(t,\hat{t}_1)$ in the Table~\ref{tab:5}. Remarkably, the data in the Table~\ref{tab:5} are also stand for the invariants derived from the D-brane system with the only D-brane parameterized by the only open moduli $\hat{t}_0$.

When the two parallel D-branes coincide, the coincident D-brane in the Calabi-Yau 3-fold corresponds to the singular F-theory 4-fold in terms of Type II/F theory duality. From the gauge theory point of view, the gauge group on the worldvolume of the D-branes is enhanced to the non-Abelian group $SU(2)$.
Table~\ref{tab:6} shows the $SU(2)$ Ooguri-Vafa invariants extracted from the coincident D-brane superpotential $\mathcal{W}_c(t,\hat{t})$.

The $U(1)$ Ooguri-Vafa invariants for the individual D-brane are different from the $SU(2)$ Ooguri-Vafa invariants for coincident D-brane, such as in Table~\ref{tab:5} the first non-zero invariant is $N_{0,0,1}=54$ while in Table~\ref{tab:6} the first non-zero invariant is $N_{0,0,2}=-9$. It should be noticed that here we reduce the number of index using the condition $n_1=n_3$ in the Table~\ref{tab:5}.
Indeed, for the individual D-brane superpotential $\mathcal{W}_1(t,\hat{t}_1)$ of the D-brane system with two parallel D-branes, there are only three independent index correspond to the two closed-string moduli $t_1,~t_2$ and one open-string moduli $\hat{t}_1$.
And it also meets the expectation for the independent number of index for the Ooguri-Vafa invariants of the single D-brane system which equals to $3=2(closed)+1(open)$.
In fact $N_{n_1,n_2,n_4}$ are rightly the Ooguri-Vafa invariants for the single D-brane system as we mentioned the D-brane system with the only D-brane parameterized by the only open moduli $\hat{t}_1$. In this terms, the comparison of Table~\ref{tab:5} and Table~\ref{tab:6} shows that the disk invariants of the coincident D-brane geometry are different from the ones obtained from the single D-brane system, although these two D-brane geometry are equivalent in terms of set theory. Similar to what we have found in the two previous models, it is an evidence of the phase transition.

The following Figures show the distribution of the $U(1)$ and $SU(2)$ Ooguri-Vafa invariants when the index $n_1,~n_2$ are fixed at various values and we take the absolute value of the Ooguri-Vafa invariants making it easy to compare. The points for invariants of two phases generate the wave-packet respectively in this model too.

\begin{figure}[H]
  \setlength{\abovecaptionskip}{-1cm}
  \setlength{\belowcaptionskip}{-0.5cm}
  \centering
  \includegraphics[width=1\textwidth]{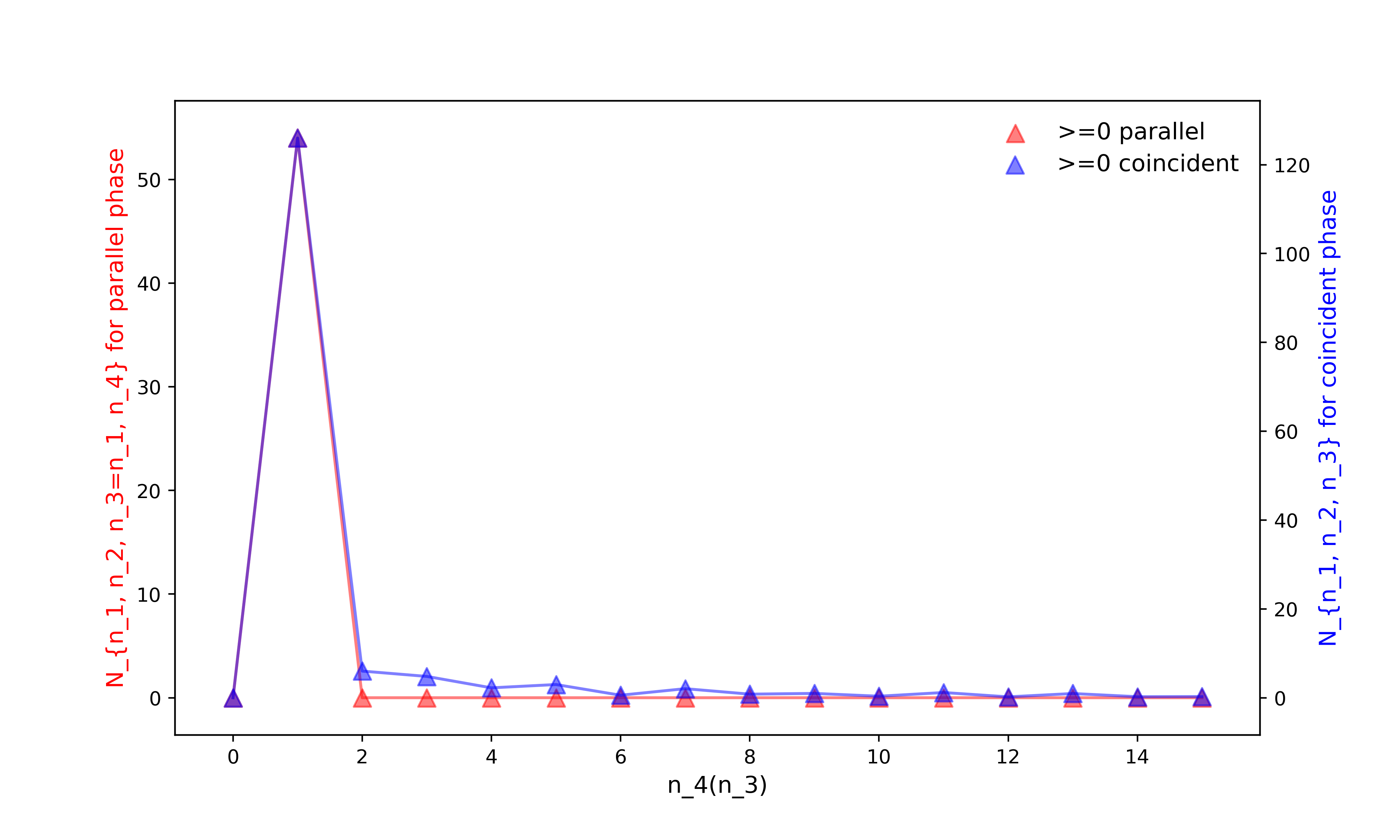}
  \caption{}\label{fig:11}
\end{figure}

In the Figure~\ref{fig:11} when $n_1=0~n_2=0$, the hight of wave-packet for the parallel and coincident phase are $54$ and $126$ respectively, while the center of the two wave-packets are coincident and the duration of them are nearly the same. Similar to the last two models, since $n_4(n_3)=2$ , $n_4$ for parallel phase and $n_3$ for coincident phase, the $U(1)$ and $SU(2)$ Ooguri-Vafa invariants tend to become zero as the increase of $n_4(n_3)$.

\begin{figure}[H]
  \setlength{\abovecaptionskip}{-1cm}
  \setlength{\belowcaptionskip}{-0.5cm}
  \centering
  \includegraphics[width=1\textwidth]{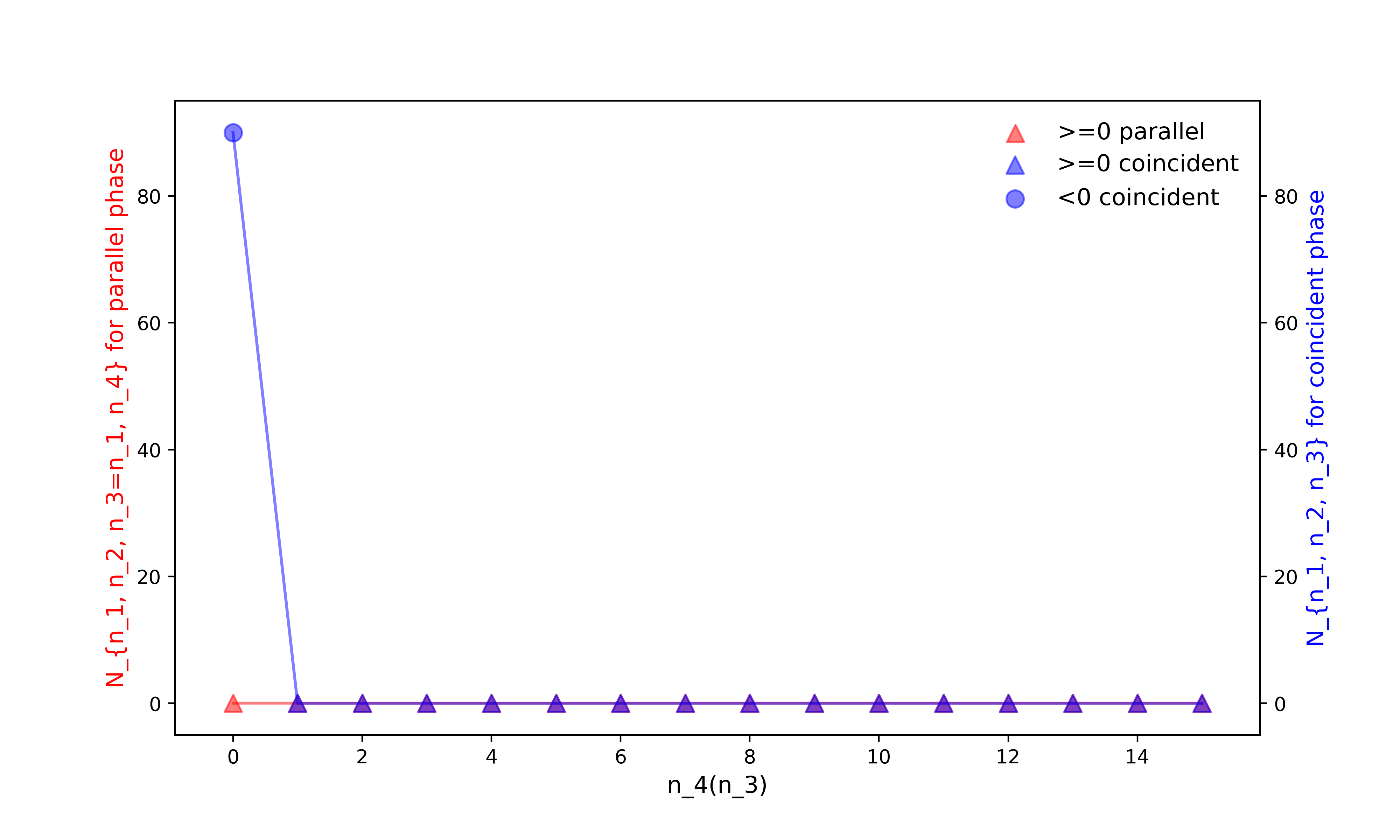}
  \caption{}\label{fig:12}
\end{figure}

In the Figure~\ref{fig:12} as $n_1=0~n_2=1$, all the points for the parallel phase are lie on the horizontal axis $N_{n_1, n_2, n_3, n_4}=0$, while the only non-zero point for the coincident phase is $N_{0,1,0}=-90$. In fact the red lines can be viewed as a wave-packet with hight zero and the blue lines form a half wave-packet.

\begin{figure}[H]
  \setlength{\abovecaptionskip}{-1cm}
  \setlength{\belowcaptionskip}{-0.5cm}
  \centering
  \includegraphics[width=1\textwidth]{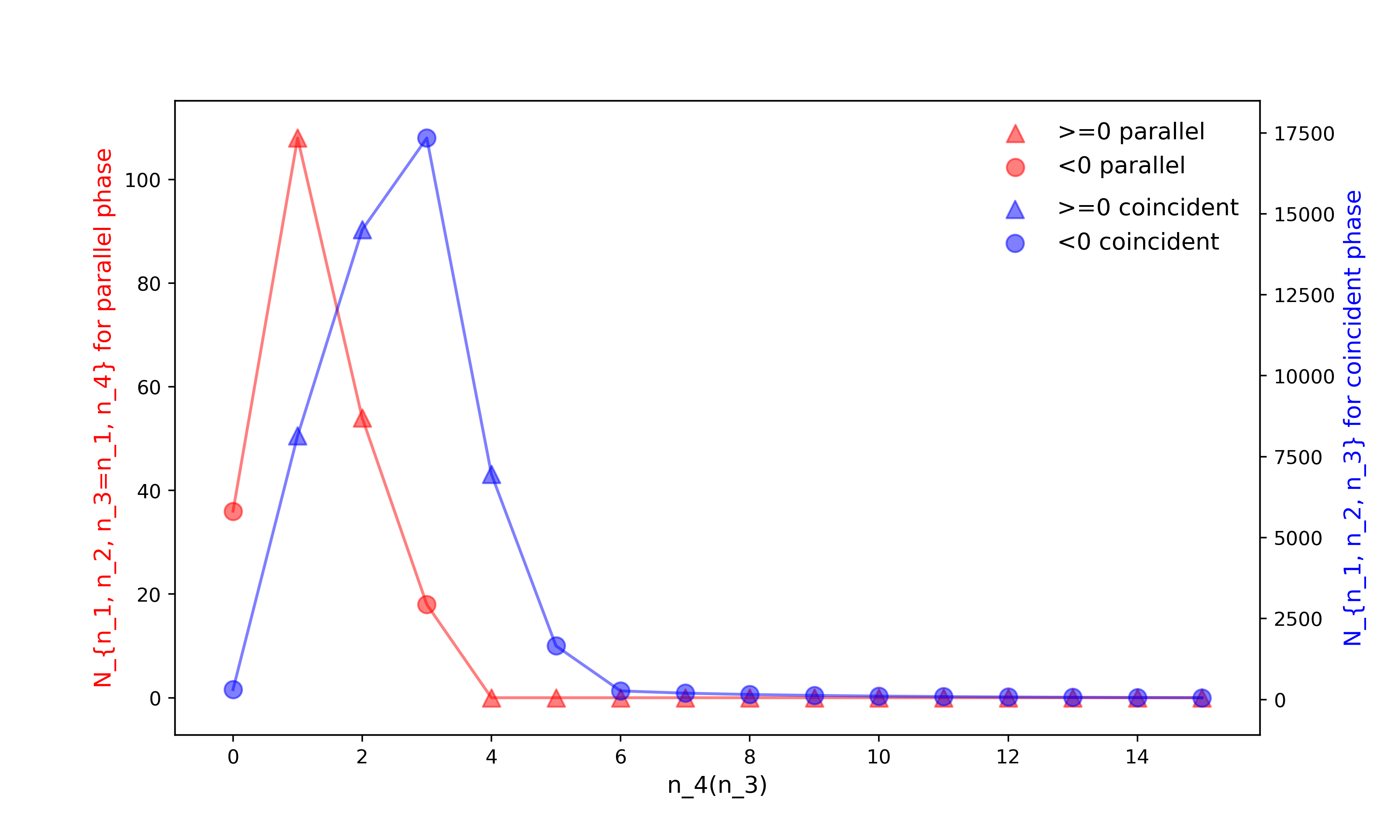}
  \caption{}\label{fig:13}
\end{figure}

In the Figure~\ref{fig:13} when $n_1=1~n_2=0$, the peak value of the Ooguri-Vafa invariants for parallel phase is $N_{1,0,1,1}=108$, while the highest point of blue is $N_{1,0,3}= -86697/5$. The center of blue wave-packet shifts right 2 unit from the center of red wave-packet and the duration of coincident one (6) is wider than the parallel one (4).

\begin{figure}[H]
  \setlength{\abovecaptionskip}{-1cm}
  \setlength{\belowcaptionskip}{-0.5cm}
  \centering
  \includegraphics[width=1\textwidth]{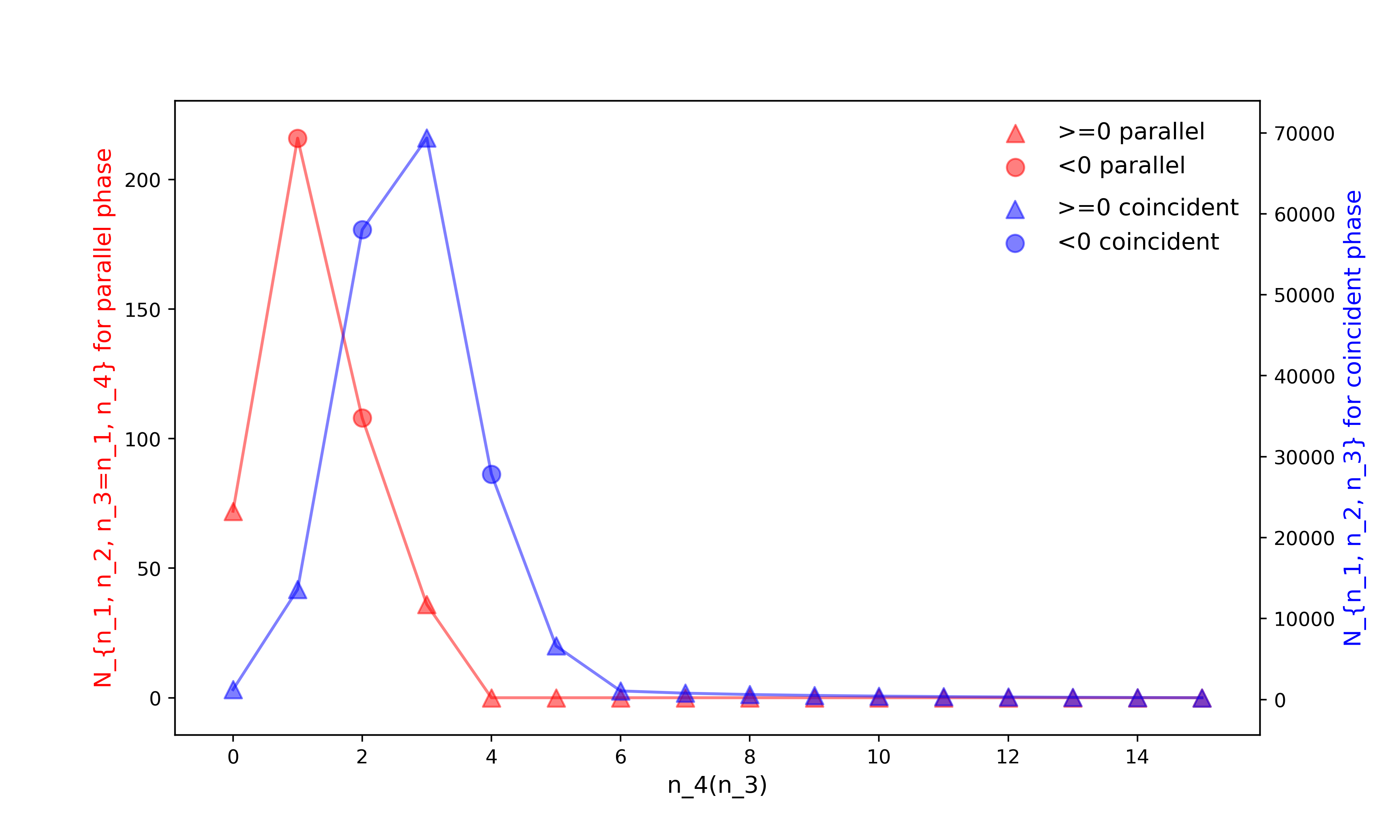}
  \caption{}\label{fig:14}
\end{figure}

In the Figure~\ref{fig:14} when $n_1=1~n_2=1$, the peak value for parallel phase is $N_{1,1,1,1}=216$, while the peak value for coincident phase is $N_{1,1,3}=346788/5$ which is over 320 times of the $N_{1,0,1,1}$.
In addition, the duration of the wave-packets for coincident phase is wider than it for parallel phase, to be precise, the durations are 6 and 4 respectively. Also we see the shift of the center of blue wave-packet comparing the center of red one obviously.

As what we find in the last two models, the wave-packets for coincident phase are higher and wider than the ones for parallel phase. With the increase of $n_1,~n_2$, the gap of the peak values between the invariants of two phases become farther, while with the increase of $n_4(n_3)$ the value of invariants for two phases are getting closer and near to zero.
Moreover the distinction of the center of wave-packet for two phases appears. These diagrams reflect the spectrum of BPS states
for the two different phases of the D-brane system. As we mentioned, the coincidence of the D-branes generates more complicate state spectrum than the situation of parallel. It suggests the appearance of the phase transition in this model.

\section{Summary and discussion}
The D-brane superpotential plays a crucial role in both physics and mathematics.
Physically speaking, they determine the vacuum of the low energy effective theory. In the view of A-model, it is the generating functional of the Ooguri-Vafa invariants of Calabi-Yau manifold involving D-branes. These Ooguri-Vafa invariants are closely related to the number of the BPS states.
Mathematically speaking, they count the holomorphic disks on Calabi-Yau manifolds. However, there is no systemical mathematical theory for these invariant up to now, the results in this paper provide a prediction of them.
Furthermore, the study of the phase structure of D-brane system is also quite meaningful since the D-brane system gives rise to the different theories in
various phases and bridges them through the phase transition.

In this paper, we study the parallel phase and the coincident phase of three different D-brane models with compact Calabi-Yau manifold compactification.
They are also known as the Coulomb branch and the Higgs branch in the language of gauge theory and the phase transition connect the $U(1)\times U(1) \times ...\times U(1)$ gauge theory and $SU(n)$ gauge theory on the worldvolume of D-branes when the $n$ parallel D-branes approach to each other and melt into one.
Using the open-closed duality, the D-brane superpotentials of different phases are obtained near the large radius limit point in the open-closed deformation space and the disk invariants are extracted from the instanton expansion of them.

In the parallel D-branes phase of all the three D-brane systems,  namely the D-branes on the mirror quintic and the hypersurface in the weighted projective space $P(1,1,2,2,2)$ and $P(1,1,1,3,3)$ respectively, the results show the discrete $Z_2$ symmetry of the superpotentials contributed by each individual D-brane.
Correspondingly, the $U(1)$ Ooguri-Vafa invariants extracted from the superpotentials of the two parallel D-branes are the same up to a transformation of the index and are in agreement with the results derived from the D-brane system which contains only one of the two parallel D-branes.
In the coincident phase, although the submanifold corresponding to the D-brane system with coincident D-branes are the same as  the submanifold corresponding to the D-brane system which contains only one D-brane in terms of the set theory, the Ooguri-Vafa invariants derived from the two systems are different in our calculation for each model. We present the first several order Ooguri-Vafa invariants with the figures and find that these points for parallel phase and the coincident phase form wave-packet both. In all three models, the wave-packets for parallel phase are higher than coincident phase. As the fixed index become bigger, the difference of the peak values becomes extremely large, while as the free index increasing, the $U(1)$ and $SU(2)$ Ooguri-Vafa invariants are getting closer and near to zero finally. Furthermore, when the two D-branes coincide the center of the wave-packets for the Ooguri-Vafa invariants seemly shift to the right side of the center of wave-packets for the parallel phase, namely the direction of the larger free index. In other words, the center of wave-packets for the two phases are at distinct locations. One more observation in our models is that generally the durations of the wave-packets for coincident phase are wider than the durations for parallel phase.
All the observations suggest that the coincidence of two parallel D-branes leads to more complicate spectrum structure of BPS states.
It can be interpreted as the evidence of the phase transition between the Coulomb branch and the Higgs branch.

Following, we will study non-abelian nature of superpotentials \cite{LMRS.05,JL.05,GMM.05,M.05} and the phase transition for more general D-brane system with multiple open-string deformation. We are going to understand in greater detail about the physical properties of the $SU(2)$ Ooguri-Vafa invariants. Furthermore, we try to explain the rationality of these $SU(2)$ invariants, find an multi-covering formula which encodes some index and search deeper relation between the $SU(2)$ Ooguri-Vafa invariants and the $SU(2)$ gauge theory.
It is also interesting to calculate the D-brane superpotential from the $A_{\infty}$-structure in the derived category of coherent sheaves of Calabi-Yau manifold and path algebras of quivers.

\acknowledgments

This work is supported by NSFC(11475178) and Y4JT01VJ01.


\begin{appendices}
\section{Tables}

\begin{table}[H]
\begin{center}
\label{table1}
\def\temptablewidth{1.0\textwidth}
\begin{tabular*}{\temptablewidth}{@{\extracolsep{\fill}}c|ccccc}
 \hline $n_2=n_3\backslash n_1$&0&1&2&3&4  \\ \hline
 0&0& -320& 13280& -1088960& 119783040 \\
 1&20& 1600& -116560&
12805120& -1766329640  \\
 2&0& 2040& 679600& -85115360&
13829775520  \\
 3&0& -1460& 1064180& 530848000& -83363259240 \\
 4&0& 520& -1497840& 887761280&
541074408000  \\ \hline
\end{tabular*}
\tabcolsep 0pt \caption{\label{tab:1} $U(1)$ Ooguri-Vafa invariants
$N_{n_1,n_2,n_3}$ for the superpotential $\mathcal{W}_1(t,\hat{t}_1)$ for one of the two parallel branes
on the mirror quintic.} \vspace*{-12pt}
\end{center}
\end{table}

\begin{table}[H]
\begin{center}
\label{table1}
\def\temptablewidth{1.0\textwidth}
\begin{tabular*}{\temptablewidth}{@{\extracolsep{\fill}}c|ccccc}
 \hline $n_2\backslash n_1$&0&1&2&3&4  \\ \hline
 0&0& -88128 & 1011258272 & $ - \frac{{115773147850176}}{5}$ & $\frac{{4945105151970217984}}{{7}}$ \\
 1& 52 & 11647296 & -260889514960 &
$\frac{{56843178601939968}}{5}$ & $ - \frac{{4004883494965708695000}}{7}$  \\
 2&$ \frac{{8}}{3}$ & 41096760 & $\frac{{52555077573328}}{3}$ & $ - \frac{{3779988263028158048}}{{5}}$ &
$\frac{{393821325944767198218688}}{{7}}$  \\
 3& $ \frac{{32}}{{15}}$ & -110965236 & 89574224256660 & $\frac{{252677495138846575872}}{5}$ & $ - \frac{{19798748028733225110625368}}{7}$ \\
 4& $ \frac{{104}}{{105}}$ & 192993672 & -536403617989440 & 291434182352136870528 &
$\frac{{28082647576600988112990258560}}{{147}}$  \\ \hline
\end{tabular*}
\tabcolsep 0pt \caption{\label{tab:2} $SU(2)$ Ooguri-Vafa invariants
$N_{n_1,n_2}$ for the of-shell superpotential $\mathcal{W}_c(t,\hat{t})$ of coincident branes
on the mirror quintic.} \vspace*{-12pt}
\end{center}
\end{table}

\begin{table}[H]
\begin{center}
\def\temptablewidth{1.0\textwidth}
\begin{tabular*}{\temptablewidth}{@{\extracolsep{\fill}}c|ccccc}
\hline $n_1=0~n_2\backslash n_3=n_4$&0&1&2&3&4  \\ \hline
 0&0& 8& 0& 0& 0 \\
 1&2& 0& 0&
0& 0  \\
 2&0& 0& 0& 0&
0  \\
 3&0& 0& 0& 0& 0  \\
 4&0& 0&0 & 0& 0  \\ \hline
\end{tabular*}
\end{center}
\begin{center}
\def\temptablewidth{1.0\textwidth}
\begin{tabular*}{\temptablewidth}{@{\extracolsep{\fill}}c|ccccc}
\hline $n_1=1~n_2\backslash n_3=n_4$&0&1&2&3&4  \\ \hline
 0&-18& 0& 44& -32& 6 \\
 1&-42& 240& 188&
-80& 14  \\
 2&0& 0& 0& 0&
0  \\
 3&0& 0& 0& 0& 0  \\
 4&0& 0&0 & 0& 0  \\ \hline
\end{tabular*}
\end{center}
\begin{center}
\def\temptablewidth{1.0\textwidth}
\begin{tabular*}{\temptablewidth}{@{\extracolsep{\fill}}c|ccccc}
\hline $n_1=2~n_2\backslash n_3=n_4$&0&1&2&3&4  \\ \hline
 0&72& -216& 0& 744& -1392 \\
 1&744& -6192& 26016&
34704& -34724  \\
 2&204& -1368& 7848& 6504&
-6024  \\
 3&0& 0& 0& 0& 0  \\
 4&0& 0&0 & 0& 0  \\ \hline
\end{tabular*}
\end{center}
\begin{center}
\def\temptablewidth{1.0\textwidth}
\begin{tabular*}{\temptablewidth}{@{\extracolsep{\fill}}c|ccccc}
\hline $n_1=3~n_2\backslash n_3=n_4$&0&1&2&3&4  \\ \hline
 0&-486& 2624& -4892& 0& 19536 \\
 1&-14136& 137232& -733644&
2675536& 4301824  \\
 2&-19032& 198576& -1077372& 5602640&
6751552  \\
 3&-1566& 11744& -55820& 341328& 278016  \\
 4&0& 0&0 & 0& 0  \\ \hline
\end{tabular*}
\tabcolsep 0pt \caption{\label{tab:3} $U(1)$ Ooguri-Vafa invariants
$N_{n_1,n_2,n_3,n_4}$ for the superpotential $\mathcal{W}_1(t,\hat{t}_1)$ for one of the two parallel branes
on the $X_8(1,1,2,2,2)$.} \vspace*{-12pt}
\end{center}
\end{table}

\begin{sidewaystable}[h]
\begin{center}
\def\temptablewidth{1.0\textwidth}
\begin{tabular*}{\temptablewidth}{@{\extracolsep{\fill}}c|ccccc}
\hline $n_1=0~n_2\backslash n_3$&0&1&2&3&4  \\ \hline
 0&0& 30& $\frac{3}{2}$ & $\frac{6}{5}$ & $\frac{39}{70}$ \\
 1&12& 0& 0&
0& 0  \\
 2&0& 0& 0& 0&
0  \\
 3&0& 0& 0& 0& 0  \\
 4&0& 0&0 & 0& 0  \\ \hline
\end{tabular*}
\end{center}
\begin{center}
\def\temptablewidth{1.0\textwidth}
\begin{tabular*}{\temptablewidth}{@{\extracolsep{\fill}}c|ccccc}
\hline $n_1=1~n_2\backslash n_3$&0&1&2&3&4  \\ \hline
 0&-1269& 49518& 228798& -441396& $\frac{{15950397}}{{35}}$ \\
 1&-7236& 551448& 1308312&
-2340144& $\frac{{84793188}}{{35}}$  \\
 2&-2889& 306990& 523638& -872316&
$\frac{{31694097}}{{35}}$  \\
 3&0& 0& 0& 0& 0  \\
 4&0& 0&0 & 0& 0  \\ \hline
\end{tabular*}
\end{center}
\begin{center}
\def\temptablewidth{1.0\textwidth}
\begin{tabular*}{\temptablewidth}{@{\extracolsep{\fill}}c|ccccc}
\hline $n_1=2~n_2\backslash n_3$&0&1&2&3&4  \\ \hline
 0&$\frac{{332037}}{2}$& -25407684& $\frac{{116159301}}{2}$& 1066650360& $ - \frac{{62988898977}}{7}$ \\
 1&4299300&-680827392& 20224177488&
88201957920& $ - \frac{{2490301459944}}{7}$  \\
 2&11684034& -1835666712& 71930694546& 241982813880&
$ - \frac{{6606010975860}}{7}$  \\
 3&6488100& -1005651072& 47567629872& 133209460320& $ - \frac{{3552739749864}}{7}$  \\
 4&$\frac{{907497}}{2}$& -66961404& $\frac{{4891301217}}{2}$ & 5101825560& $ - \frac{{178876334277}}{7}$  \\ \hline
\end{tabular*}
\end{center}
\begin{center}
\def\temptablewidth{1.0\textwidth}
\begin{tabular*}{\temptablewidth}{@{\extracolsep{\fill}}c|ccccc}
\hline $n_1=3~n_2\backslash n_3 $&0&1&2&3&4  \\ \hline
 0&-40214583& 14628280986& -75342707334& $ - \frac{{1166211861138}}{{25}}$& $\frac{{160918023374856}}{{35}}$ \\
 1& -2543758128 & 782000827848& -25978588599240&
$\frac{{15775718348546352}}{{25}}$ & $\frac{{120563651255232096}}{{35}}$  \\
 2&-19688760180& 5866912290330& -232396059792654& $\frac{{38538316316561364}}{5}$&
$\frac{{237577376428491240}}{7}$ \\
 3&-40157723916& 11853517750560& -478953286932672& 18828098652125736& $\frac{{496479648923217360}}{7}$  \\
 4&-25972637580& 7619541285330&-304637338603494 & $\frac{{67806211523055876}}{5}$& $\frac{{314656357375628640}}{7}$  \\ \hline
\end{tabular*}
\tabcolsep 0pt \caption{\label{tab:4} $SU(2)$ Ooguri-Vafa invariants
$N_{n_1,n_2,n_3}$ for the superpotential $\mathcal{W}_c(t,\hat{t})$ for coincident brane
on the $X_8(1,1,2,2,2)$.} \vspace*{-12pt}
\end{center}
\end{sidewaystable}

\begin{table}[h]
\begin{center}
\def\temptablewidth{1.0\textwidth}
\begin{tabular*}{\temptablewidth}{@{\extracolsep{\fill}}c|ccccc}
\hline $n_1=n_3=0,~n_2\backslash n_4$&0&1&2&3&4  \\ \hline
 0&0& 54& 0& 0& 0 \\
 1&9& 0& 0&
0& 0  \\
 2& 0& 0& 0& 0&
0  \\
 3& 0& 0& 0& 0& 0  \\
 4& 0& 0&0 & 0& 0  \\ \hline
\end{tabular*}
\end{center}
\begin{center}
\def\temptablewidth{1.0\textwidth}
\begin{tabular*}{\temptablewidth}{@{\extracolsep{\fill}}c|ccccc}
\hline $n_1=n_3=1,~n_2\backslash n_4$&0&1&2&3&4  \\ \hline
 0&-36& 108& 54& -18& 0 \\
 1&72& -216& -108& 36& 0    \\
 2&-180& 540& 270& -90&
0  \\
 3& 1152& -3456& -1728& 576& 0  \\
 4& -10296& 30888& 15444& -5148& 0  \\ \hline
\end{tabular*}
\end{center}
\begin{center}
\def\temptablewidth{1.0\textwidth}
\begin{tabular*}{\temptablewidth}{@{\extracolsep{\fill}}c|ccccc}
\hline $n_1=n_3=2,~n_2\backslash n_4$&0&1&2&3&4  \\ \hline
 0& 18& -54& 216& 36& 0 \\
 1& -36& -1728& 4860& 2772& -1026 \\
 2& 108& 7020& -19872& -11160& 4104  \\
 3& -756& -63504& 183708& 99036& -35154  \\
 4& 6570& 738450& -2156760& -1140300& 398520  \\ \hline
\end{tabular*}
\end{center}
\begin{center}
\def\temptablewidth{1.0\textwidth}
\begin{tabular*}{\temptablewidth}{@{\extracolsep{\fill}}c|ccccc}
\hline $n_1=n_3=3,~n_2\backslash n_4$&0&1&2&3&4  \\ \hline
 0& 0& 0& -54& 432& 54 \\
 1& -1224& 17280& -80460& 203256& 243756  \\
 2& -108& -5832& -97686& 248184& 174960  \\
 3& 648& 46656& 1408536& -3851928& -2341656  \\
 4& -5508& -507384& -22617954& 63675720& 36426672  \\ \hline
\end{tabular*}
\tabcolsep 0pt \caption{\label{tab:5} $U(1)$ Ooguri-Vafa invariants
$N_{n_1,n_2,n_3,n_4}$ for the superpotential $\mathcal{W}_1(t,\hat{t})$ for one of the two parallel branes
on the $X_8(1,1,1,3,3)$.} \vspace*{-12pt}
\end{center}
\end{table}

\begin{table}[h]
\begin{center}
\def\temptablewidth{1.0\textwidth}
\begin{tabular*}{\temptablewidth}{@{\extracolsep{\fill}}c|ccccc}
\hline $n_1=0~n_2\backslash n_3$&0&1&2&3&4  \\ \hline
 0& 0& 126& 6& $\frac{24}{5}$& $\frac{78}{35}$ \\
 1& -90& 0& 0&
0& 0  \\
 2& 468& 0& 0& 0&
0  \\
 3& -3726& 0& 0& 0& 0  \\
 4& 39024& 0& 0& 0& 0  \\ \hline
\end{tabular*}
\end{center}
\begin{center}
\def\temptablewidth{1.0\textwidth}
\begin{tabular*}{\temptablewidth}{@{\extracolsep{\fill}}c|ccccc}
\hline $n_1=1~n_2\backslash n_3$&0&1&2&3&4  \\ \hline
 0& -306& 8145& 14508& $-\frac{86697}{5}$& $\frac{243648}{35}$ \\
 1& 1224& 13590& -58032& $\frac{346788}{5}$& $-\frac{974592}{35}$  \\
 2& -5742& -343296& 283302& $-\frac{1523583}{5}$& $\frac{633096}{5}$ \\
 3& 33480& 5447682& -1646352& $\frac{8934948}{5}$& $-\frac{3701376}{5}$  \\
 4& -285030& -79672185& 13993200& -15256215& $\frac{44174160}{7}$  \\ \hline
\end{tabular*}
\end{center}
\begin{center}
\def\temptablewidth{1.0\textwidth}
\begin{tabular*}{\temptablewidth}{@{\extracolsep{\fill}}c|ccccc}
\hline $n_1=2~n_2\backslash n_3$&0&1&2&3&4  \\ \hline
 0& 513& -70704& -925161& $-\frac{15798834}{5}$& $\frac{118139829}{35}$ \\
 1& 58824& -6041160& 151385790& $\frac{1727841672}{5}$& $-\frac{21669017772}{35}$  \\
 2& -137160& 19345032& -320285505& $-\frac{6367696146}{5}$& $\frac{67703910861}{35}$  \\
 3& 291168& -88498944& -2697837168& 6473257344& $ -8441996412$  \\
 4& -1614393& 814794210& 76558143723& $-\frac{287812981512}{5}$& $\frac{2507028719457}{35}$ \\ \hline
\end{tabular*}
\end{center}
\begin{center}
\def\temptablewidth{1.0\textwidth}
\begin{tabular*}{\temptablewidth}{@{\extracolsep{\fill}}c|ccccc}
\hline $n_1=3~n_2\backslash n_3 $&0&1&2&3&4  \\ \hline
 0& $-\frac{4752}{5}$& $\frac{4774464}{5}$& $\frac{123738138}{5}$& $\frac{4393652358}{25}$& $\frac{24739662618}{35}$ \\
 1& $-\frac{8864856}{5}$ & $\frac{1384883352}{5}$& $-\frac{33880795776}{5}$&
$\frac{1212911728794}{25}$ & $\frac{5078618177904}{35}$  \\
 2& 8733798& -1302872688& 18528872472& $\frac{2324479046526}{5}$&
$\frac{4029114362220}{7}$ \\
 3& $-\frac{263030976}{5}$& $\frac{37059726312}{5}$ & $-\frac{488076797016}{5}$& $-\frac{47266859201286}{25}$& $-\frac{177279589143936}{35}$  \\
 4& $\frac{1617306858}{5}$& $-\frac{220114519416}{5}$& $\frac{807888819618}{5}$ & $-\frac{407658185581302}{25}$ & $\frac{2639629362439008}{35}$  \\ \hline
\end{tabular*}
\tabcolsep 0pt \caption{\label{tab:6} $SU(2)$ Ooguri-Vafa invariants
$N_{n_1,n_2,n_3}$ for the superpotential $\mathcal{W}_c(t,\hat{t})$ for coincident brane
on the $X_8(1,1,1,3,3)$.} \vspace*{-12pt}
\end{center}
\end{table}

\end{appendices}

\FloatBarrier

\end{document}